\documentclass[aps, prd, preprint, preprintnumbers, showpacs,
showkeys, amsfonts, nofootinbib, floatfix,10]{revtex4-1}
\usepackage{graphicx}
\usepackage{sparticles}
\usepackage{epsfig}
\usepackage{rotate}
\usepackage{rotating}
\usepackage{multirow}
\usepackage{longtable}
\usepackage[T1]{fontenc}
\usepackage{array}
\usepackage{graphicx,amssymb,amsfonts,amsmath}

\def\fo{\hbox{{1}\kern-.
25em\hbox{l}}}

\def\beq{\begin{equation}}
\def\eeq{\end{equation}}
\def\eq{\end{equation}}
\def\to{\rightarrow}

\def\bsg{\ifmmode B\to X_s\gamma\else $B\to X_s\gamma$\fi}
\def\bsll{\ifmmode B\to X_s\ell^+\ell^-\else $B\to X_s\ell^+\ell^-$\fi}
\def\bstt{\ifmmode B\to X_s\tau^+\tau^-\else $B\to X_s\tau^+\tau^-$\fi}
\def\shat{\ifmmode \hat{s}\else $\hat{s}$\fi}

\def\EmissT{\not \! \!  E_{T}}

\def\s2b{s_{2\beta}}

\def\u1{G_{SM}\otimes U(1)^{\prime}}

\newcommand{\newc}{\newcommand}
\newcommand{\be}{\begin{equation}}
\newcommand{\ee}{\end{equation}}

\newcommand{\bea}{\begin{eqnarray}}
\newcommand{\eea}{\end{eqnarray}}

\newc{\lcal}{\int {\cal L}dt}

\newc{\lsp}{{\widetilde{\chi}^0_1}}
\newc{\niki}{{\widetilde{\chi}^0_2}}
\newc{\nuc}{{\widetilde{\chi}^0_3}}
\newc{\ndort}{{\widetilde{\chi}^0_4}}
\newc{\nbes}{{\widetilde{\chi}^0_5}}
\newc{\nalti}{{\widetilde{\chi}^0_6}}
\newc{\mnbir}{{M_{\widetilde{\chi}^0_1}}}
\newc{\mniki}{{M_{\widetilde{\chi}^0_2}}}
\newc{\mnuc}{{M_{\widetilde{\chi}^0_3}}}
\newc{\mndort}{{M_{\widetilde{\chi}^0_4}}}
\newc{\mnbes}{{M_{\widetilde{\chi}^0_5}}}
\newc{\mnalti}{{M_{\widetilde{\chi}^0_6}}}
\newc{\stauR}{{\widetilde{\tau}_R}}
\newc{\stau}{{\widetilde{\tau}_1}}
\newc{\staut}{{\widetilde{\tau}_2}}
\newc{\mstop}{m_{\widetilde{t}}}
\newc{\mHpm}{m_{H^\pm}}
\newc{\simgt}{\lower.7ex\hbox{$\;\stackrel{\textstyle>}{\sim}\;$}}
\newc{\simlt}{\lower.7ex\hbox{$\;\stackrel{\textstyle<}{\sim}\;$}}
\newc{\ie}{{\it i.e.}}
\newc{\etal}{{\it et al.}}
\newc{\eg}{{\it e.g.}}
\newc{\kev}{\hbox{\rm\,keV}}
\newc{\mev}{\hbox{\rm\,MeV}}
\newc{\gev}{\hbox{\rm\,GeV}}
\newc{\tev}{\hbox{\rm\,TeV}}
\newc{\xpb}{\hbox{\rm\, pb}}
\newc{\xfb}{\hbox{\rm\, fb}}

\newc{\mtop}{m_t}
\newc{\mbot}{m_b}
\newc{\mz}{m_Z}
\newc{\mw}{M_W}
\newc{\alphasmz}{\alpha_s(m_Z^2)}
\newc{\swsq}{\sin^2\theta_W}
\newc{\tw}{\tan\theta_W}
\newc{\cw}{\cos\theta_W}
\newc{\sw}{\sin\theta_W}
\newc{\BR}{\hbox{\rm BR}}
\newc{\zbb}{Z\to b\bar}
\newc{\Gb}{\Gamma (Z\to b\bar b)}
\newc{\Gh}{\Gamma (Z\to \hbox{\rm hadrons})}
\newc{\rbsm}{R_b^\hbox{\rm sm}}
\newc{\rbsusy}{R_b^\hbox{\rm susy}}
\newc{\drb}{\delta R_b}

\newc{\sgn}{\mbox{sgn}}
\newc{\tbeta}{\tan\beta}
\newc{\uL}{{\widetilde{u}_L}}
\newc{\uR}{{\widetilde{u}_R}}
\newc{\cL}{{\widetilde{c}_L}}
\newc{\cR}{{\widetilde{c}_R}}
\newc{\tL}{{\widetilde{t}_L}}
\newc{\tR}{{\widetilde{t}_R}}
\newc{\dL}{{\widetilde{d}_L}}
\newc{\dR}{{\widetilde{d}_R}}
\newc{\sL}{{\widetilde{s}_L}}
\newc{\sR}{{\widetilde{s}_R}}
\newc{\bL}{{\widetilde{b}_L}}
\newc{\bR}{{\widetilde{b}_R}}
\newc{\eL}{{\widetilde{e}_L}}
\newc{\eR}{{\widetilde{e}_R}}
\newc{\mhp}{m_{H^\pm}}
\newc{\mhalf}{m_{1/2}}
\newc{\emt}{{e/\mu /\tau}}

\newc{\lR}{\widetilde{\ell}_R}
\newc{\lL}{\widetilde{\ell}_L}
\newc{\nL}{\widetilde{\nu}_{\ell_L}}
\newc{\nR}{\widetilde{\nu}_{\ell_R}}
\newc{\neL}{\widetilde{\nu}_{e_L}}
\newc{\nmL}{\widetilde{\nu}_{\mu_L}}
\newc{\nlL}{\widetilde{\nu}_{\tau_L}}
\newc{\neR}{\widetilde{\nu}_{e_R}}
\newc{\nmR}{\widetilde{\nu}_{\mu_R}}
\newc{\nlR}{\widetilde{\nu}_{\tau_R}}
\newc{\naa}{\widetilde{\chi}^0_1}
\newc{\nbb}{\widetilde{\chi}^0_2}
\newc{\ncc}{\widetilde{\chi}^0_3}
\newc{\ndd}{\widetilde{\chi}^0_4}
\newc{\nee}{\widetilde{\chi}^0_5}
\newc{\nff}{\widetilde{\chi}^0_6}
\newc{\caa}{\widetilde{\chi}^{\pm}_1}
\newc{\cbb}{\widetilde{\chi}^{\pm}_2}
\newc{\phit}{\phi_t}
\newc{\phib}{\phi_b}
\newc{\phiew}{\phi_{ew}}
\newc{\htz}{h^0_t}
\newc{\hbz}{h^0_b}
\newc{\hewz}{h^0_{ew}}
\newc{\hsmz}{h^0_{sm}}
\newc{\huz}{h^0_u}
\newc{\hsusyz}{h^0_{susy}}
\newc{\lmop}{\rm LM1^\prime}
\newc{\lmtp}{\rm LM2^\prime}
\newc{\lmsp}{\rm LM6^\prime}
\newc{\smin}{\hat{s}_{\rm min}^{1/2}}

\def\mp{M_{Pl}}

\def\lsim{\raise0.3ex\hbox{$<$\kern-0.75em\raise-1.1ex\hbox{$\sim$}}}
\def\gsim{\raise0.3ex\hbox{$>$\kern-0.75em\raise-1.1ex\hbox{$\sim$}}}
\def\to{\rightarrow} \def\ie{{\it i.e. }} \def\nn{\nonumber}
 \def\vev#1{\langle #1 \rangle} \def\gev{~{\rm
GeV}} \def\tev{~{\rm TeV}}   
   
  \def\eK{\varepsilon_K}

\def\slashchar#1{\setbox0=\hbox{$#1$}           % set a box for #1
   \dimen0=\wd0                                 % and get its size
   \setbox1=\hbox{/} \dimen1=\wd1               % get size of /
   \ifdim\dimen0>\dimen1                        % #1 is bigger
      \rlap{\hbox to \dimen0{\hfil/\hfil}}      % so center / in box
      #1                                        % and print #1
   \else                                        % / is bigger
      \rlap{\hbox to \dimen1{\hfil$#1$\hfil}}   % so center #1
      /                                         % and print /
   \fi}                                         %
\catcode`@=11
\long\def\@caption#1[#2]#3{\par\addcontentsline{\csname
  ext@#1\endcsname}{#1}{\protect\numberline{\csname
  the#1\endcsname}{\ignorespaces #2}}\begingroup
    \small
    \@parboxrestore
    \@makecaption{\csname fnum@#1\endcsname}{\ignorespaces #3}\par
  \endgroup}
\catcode`@=12

\begin{document}

\preprint{ CUMQ/HEP 169}
\vspace{1.2cm}
\title{\Large  Higgs Bosons in supersymmetric $U(1)^\prime$ models with CP
Violation}
\author{Mariana Frank$^{(1)}$\footnote{mariana.frank@concordia.ca}}
\author{Levent Selbuz$^{(1,2)}$\footnote{levent.selbuz@eng.ankara.edu.tr}}
\author{Levent Solmaz$^{(3)}$\footnote{lsolmaz@balikesir.edu.tr}}
\author{Ismail Turan$^{(4)}$\footnote{ituran@metu.edu.tr}}

\affiliation{$^{(1)}$Department of Physics, Concordia University, 7141
Sherbrooke St. West, Montreal, Quebec, Canada H4B 1R6,}
\affiliation{$^{(2)}$Department of Engineering Physics, Ankara
University, TR06100 Ankara, Turkey,}
\affiliation{$^{(3)}$Department of Physics, Bal{\i}kesir University, TR10145, Bal{\i}kesir, Turkey,}
\affiliation{$^{(4)}$Department of Physics, Middle East Technical University, TR06531 Ankara,
Turkey.}

\date{\today}
\begin{abstract}
We study the Higgs sector of the U(1)$^\prime$-extended MSSM
with CP violation. This is an extension of the MSSM Higgs sector by one singlet field, introduced to generate the $\mu$ term dynamically.  We are particularly interested in  non-standard decays of
Higgs particles, especially of the lightest one, in the presence of CP violating phases for $\mu_{eff}$ and the soft parameters. We
present analytical expressions for  neutral and  charged
Higgs bosons masses at tree and one-loop levels, including contributions from   top and bottom scalar quark sectors. We then study the production and decay channels of the neutral Higgs for a set of benchmark points consistent with low energy data and relic density constraints. Numerical simulations show
 that a Higgs boson lighter than $2m_W$ can
decay in a
quite distinctive manner,  including invisible modes into two neutralinos ($h\to\tilde{\chi}^0\tilde{\chi}^0$) up to
$\sim 50\%$ of the time, when  kinematically allowed.  The branching ratio into  $h\to\bar{b}b$, the dominant decay in the SM, is reduced in some $U(1)^\prime$ models and enhanced in others, while the branching ratios for the decays $h \to \tau^+ \tau^-$,  $h \to WW^*$ and  $h \to Z Z^*\to 4\ell $ are always reduced with respect to their  SM expectations.
 This possibility has important implications for testing the $U(1)^\prime$ model both at the LHC and later at the ILC.
\end{abstract}
%%%%%%%%%%%%%%
%%%%%%%%%%%%%%
%\vspace*{-0.9cm}
\maketitle
%%%%%%%%%%%%%%

%%%%%%%%%%%%%%%%%%%%%%%%%%%%%%%%%%%%%%%%%%%%%%%%%%%%%%%%%%%%%%%

%\tableofcontents

\section{Introduction and Motivation}

Confirmation of the Higgs mechanism of the Standard Model (SM) of
particle physics demands discovery of the elusive Higgs boson, likely seen at ATLAS \cite{:2012gk} and CMS \cite{:2012gu} at a mass around $126$ GeV. The
Minimal Supersymmetric extension of the SM (MSSM), which is
arguably the best motivated extension of the SM, offers
stabilization of the Higgs mass, and moreover agrees well with the
SM predictions in certain portions of its restricted parameter space. For
instance, for the upper limit of  $m_h\sim$135 GeV of the MSSM
$h\to\bar{b}b$ is the dominant decay mechanism ($\sim 60\%$) in the
SM and in the MSSM. On the other hand, in gauge and Higgs extended
supersymmetric models, the properties of the Higgs bosons can be
substantially different from that of the standard supersymmetric
model predictions.  For instance, the addition of one singlet field to the MSSM Higgs sector provides new tree-level contributions to the $F$- and $D$-terms, which stabilize the Higgs mass naturally at a larger value \cite{Ross:2012nr}.
While many models predict a light Higgs boson
around the weak scale (say $\sim$100 GeV), it
will take some time to differentiate whether the boson discovered at the LHC belongs to the
SM gauge symmetry,  its minimal supersymmetric version (MSSM),
or even to another extension such as the gauge extended versions
of the MSSM.

Extensions of the  gauge symmetry by an extra $U(1)$ factor  (supersymmetric
or not) are arguably the simplest extensions of the minimal model. The best justification for these extended models arises from assuming grand unified theories of strong and electromagnetic interactions (GUTs). In GUT symmetries, it seems difficult to break most  scenarios directly to $SU(2)_L \times U(1)_Y$, as most models such as $SU(5), ~SO(10)$, or $E_6$ involve an additional $U(1)$ group in the breaking. In supersymmetric U(1)$^\prime$ models  \cite{Demir:2005ti} (referred to as U(1)$^\prime$ models from now on),  the number of the neutral Higgs bosons  is
increased by an
 additional singlet field (S) over that of the MSSM, and the vacuum expectation value (VEV) of the singlet $\langle S \rangle$ is responsible for the generation of
the $\mu$ term, which allows Higgs fields to couple to each other  \cite{Cvetic:1997ky,muprob};  while number of charged Higgs bosons in the U(1)$^\prime$ extended models remains the same  as in the MSSM.
The interest in the  Higgs sector of the U(1)$^\prime$ models also comes from the fact that
 such models arise naturally from string inspired models \cite{Cvetic:1995rj,Babu:1996vt, Cvetic:1996mf, Cvetic:1997ky}, or as  the dynamical solution to the $\mu$
problem in gauge-mediated supersymmetry breaking (GMSB) \cite{Langacker:1999hs}. While in the MSSM and in the U(1)$^\prime$ models lightest
neutralino is the best candidate for a LSP, for the latter the LSP is
less constrained.

In these models,
the lightest Higgs boson could potentially behave differently from the SM
or the MSSM Higgs boson due to its singlet nature. While a Higgs boson of mass $m_h\sim$ 126 GeV can be predicted by the SM, or by the MSSM,  or by
numerous other models, the coupling of the Higgs
to the known fermions or bosons is not the same
in all these models. This fact can be extrapolated not only from
the number of the Higgs bosons but also from their production and
decay mechanisms.

Of all the Higgs bosons in a model, the properties of the lightest neutral state are the most
interesting, also given its likely  discovery already at the LHC. An
interesting possibility is that its decay could be partially into
invisible modes (a possibility hinted at by the reduced branching ratios into fermions at the LHC), or that there is another Higgs boson lighter than the one at $126$ GeV, which decays  completely or
almost  so, invisibly \cite{Godbole:2003it,Davoudiasl:2004aj}.   This scenario is motivated by global fits to the  data at the LHC which indicate that a Higgs boson branching ratio of 64\% is still unaccounted for \cite{Espinosa:2012vu}.

In SM the Higgs can decay invisibly only into neutrinos, and this branching ratio is $\le 0.1\%$ \cite{Denner:2011mq}.  A light Higgs boson with substantial branching ratio into
invisible channels can occur in a variety of models including scenarios with light
neutralinos, spontaneously broken lepton number, radiatively
generated neutrino masses, additional singlet scalar(s) and/or right
handed neutrinos in the extra dimensions of TeV scale gravity. Among
these possibilities, invisible decay of the lightest Higgs into light
neutralinos is interesting since the light neutralinos are well
motivated candidates for the Lightest Supersymmetric Particle (LSP), providing viable relic density explanations\footnote{Note that in $U(1)^\prime$ models the LSP can be the singlino \cite{Nakamura:2006ht,Suematsu:2005bc}. }. Decays into light neutralinos are possible in models with non-universal couplings, where LEP limits can be circumvented \cite{particledata}, and in models with a light dark matter candidate.
For instance, a study
\cite{Hall:2011au} indicates that this is a possibility in $E_6$, where the lightest Higgs boson of
the Exceptional Supersymmetric Standard Model E$_6$SSM can decay
into the lightest neutralino pairs more than 95$\%$ of the time  \cite{Hall:2011zq}.

Additionally, the Higgs sector in extended models could provide  potential sources of CP violation beyond the phase
of the CKM matrix,  also important for the observed baryon
asymmetry of the universe. These phases can affect the masses and couplings of the Higgs bosons to the gauge and matter fields of the model, as was shown in studies of Higgs sectors of the MSSM \cite{Frank:2006yh} and next-to minimal supersymmetric models (NMSSM) \cite{Graf:2012hh}. The phases can also affect production and decay rates patterns, as we will show in this study. In this work, we analyze the mass spectra of all the Higgs bosons, and the production and decay rates (visible and
invisible) for the lightest Higgs in the $U(1)^{\prime}$ extended
form of the MSSM with CP-violating phases. The masses of  Higgs bosons in the $U(1)^\prime$ with CP violating phases model have received attention previously  \cite{Demir:2003ke}, but we include them here, for consistency with the determination of their decay properties. Thus, we re-visit the Higgs sector of $ U(1)^\prime$ models and calculate the masses, and in doing this, we improve on the previous calculation by including
contributions from both (s)top and (s)bottom sectors at
one-loop level, and add the constraint that the lightest neutral state should have mass $\sim$ 125 GeV.

Motivated by the above considerations,  we study  anomaly-free U(1)$^\prime$
models to probe their peculiar Higgs sector consistent with the
known  (astrophysical and collider) bounds, which are included in our benchmark points.  We add the scalar quarks and neutralino contributions, and calculate a complete spectrum for the latter, and insure agreement with the relic density, assuming that the lightest neutralino is the LPS . We then
study the production and decay modes of the lightest neutral Higgs boson,  with the purpose of unraveling  the existence and consequences of invisibly
decaying Higgs bosons within the $U(1)^{\prime}$ model.

The outline of our study is as follows. In the following section (Section \ref{sec:model}) we
 introduce our effective U(1)$^\prime$ model, with particular emphasis on the Higgs sector. We present tree-level ( \ref{subsec:higgs}) and one loop mass evaluations (\ref{subsec:loopHiggs}), and then an analytical calculation of the charged and neutral Higgs masses (\ref{subsec:massHiggs}).  We then introduce the neutralino spectrum  (\ref{subsec:neutralinos}) of the $U(1)^\prime$ model, which contains two additional neutralinos from the MSSM.  We include the constraints on the particle spectrum coming from low-energy measurements of CP violation in Section \ref{sec:constraints}, in particular from electric dipole moments (\ref{subsec:EDM}) and $\varepsilon_K$ (\ref{subsec:epsK}). Following the exposition of the model and its constraints, we present our numerical
investigations in Section \ref{sec:numerical}, in particular for the lightest neutral Higgs boson production and decay in \ref{subsec:H1}, comment on the second lightest neutral state in \ref{subsec:H2}. We  summarize our findings and conclude in Section \ref{sec:conclusion}. The full form of analytical
solutions for the masses can be found in the Appendices.

%%%%%%%%%%%%%%%%%%%%%%%%%%%%%%%%%%%%%%%%%%%%%%%%%%%%%%
\section{The $U(1)^{\prime}$ model with CP violation \label{sec:model}}
%%%%%%%%%%%%%%%%%%%%%%%%%%%%%%%%%%%%%%%%%%%%%%%%%%%%%%%

We review here briefly the $U(1)^\prime$ model, with particular emphasis to the Higgs and the neutralino sectors, as these are relevant to our study.
The superpotential for the effective U(1)$^\prime$
model is
\begin{equation}
\label{superpot} W=Y_{S}\widehat{S}
\widehat{H}_{u}\cdot\widehat{H}_{d} + Y_t
\widehat{U}^c\widehat{Q}\cdot\widehat{H}_u+ Y_b
\widehat{D}^c\widehat{Q}\cdot\widehat{H}_d,
\end{equation}
where we assumed that all Yukawa couplings except for $Y_t$ and $Y_b$ are negligible.
%Gauge invariance of $W$ under U(1)$^\prime$ requires
%$Q_u+Q_d+Q_{S}= 0$, $Q_{Q}+Q_{U}+Q_u=0$ and $Q_{Q}+Q_{D}+Q_d=0$.
As
can be seen from (\ref{superpot}), by replacing the $\mu$ parameter
with a singlet scalar (S) and a Yukawa coupling ($Y_S$), we resolved
the $\mu$ problem of the MSSM \cite{muprob}; $\mu$ is generated dynamically through the VEV of the $S$ field (see \ref{subsec:higgs}) and is expected to be of order of the weak scale.

In addition to the superpotential, the Lagrangian includes soft supersymmetry breaking terms containing  additional terms with respect to
the MSSM,  coming from
gaugino masses $M_a$ ($a=1,1',2,3$) and trilinear couplings $A_S$,
$A_t$ and $A_b$ as given below
\begin{eqnarray}
\label{soft} -{\cal L}_{soft}&=&(\sum_{a}M_a\lambda_a\lambda_a+
   A_SY_{S} S H_{u}\cdot H_{d}+
A_t Y_t \widetilde{U}^c \widetilde{Q}\cdot H_u+A_b Y_b \widetilde{D}^c \widetilde{Q}\cdot H_d+h.c.)\\
&+& m_{u}^{2}|H_{u}|^2 +
m_{d}^{2}|H_{d}|^2+m_{s}^{2}|S|^2+M_{\widetilde{Q}}^2|\widetilde{Q}|^2+
M_{\widetilde{U}}^2|\widetilde{U}|^2 +
M_{\widetilde{D}}^2|\widetilde{D}|^2+
M_{\widetilde{E}}^2|\widetilde{E}|^2+M_{\widetilde{L}}^2|\widetilde{L}|^2\,\nonumber.
\end{eqnarray}
Using Renormalization Group Equations (RGEs) these
soft SUSY breaking parameters are generically non-universal at low
energies. However, in our numerical studies, we choose not to deal with the evolution
of the RGEs and instead assign them values which do not
contradict with the current collider bounds. As we
are interested in CP violation, we assume some
of the soft breaking terms to be complex,  selected as the
trilinear terms ($A_{t,b,S}$) and the VEV of the Higgs field
 $S$, as these assignments do not conflict with present low energy data.

\subsection{The Higgs Sector at Tree-level}
\label{subsec:higgs}

The effective U(1)$^\prime$ model inherits  two  Higgs doublets
$H_u$, $H_d$ from the MSSM, and has an additional singlet field $S$, all of which can be expanded around their VEVs as
\begin{eqnarray}
\label{higgsexp} \langle H_u \rangle &=&
\frac{e^{i\theta_u}}{\sqrt{2}}\left(\begin{array}{c} \sqrt{2} H_u^{+}\\
v_u + \phi_u + i \varphi_u\end{array}\right) \:,\:\:\:\: \langle H_d
\rangle = \frac{e^{i\theta_d}}{\sqrt{2}}\left(\begin{array}{c} v_d +
\phi_d + i \varphi_d\\ \sqrt{2} H_d^{-}\end{array}\right), \nonumber\\
\langle S \rangle &=& \frac{e^{i\theta_s}}{\sqrt{2}} \left(v_S +
\phi_S + i \varphi_S\right),
\end{eqnarray}
in which $v^2\equiv v_u^2+v_d^2=(246\,{\rm GeV})^2$. The fields in the superpotential are charged under the $U(1)^\prime$ gauge group with charges ${\cal Q}$, required by gauge invariance to satisfy:
$$ {\cal Q}_{H_u} +{\cal Q}_{H_d}+ {\cal Q}_S=0,  \qquad {\cal Q}_{Q_3}+{\cal Q}_{U_3}+ {\cal Q}_{H_u}=0, \qquad {\cal Q}_{Q_3}+{\cal Q}_{D_3}+ {\cal Q}_{H_d}=0.  $$
  The effective $\mu$ parameter
is generated by the singlet VEV $\langle S \rangle$,  defined
as
\begin{eqnarray}
\label{eff:mu-term}
 \mu_{eff} \equiv \mu \, e^{i \theta_s},\text{ where }
\mu=\frac{Y_S v_S}{\sqrt{2}},
\end{eqnarray}
so that with this convention $\mu$ is always real. For the remaining
parameters we adopt the convention that the parameters are real, and explicitly attach CP violating phases where needed. Explicitly,
$\arg(A_t)=\theta_t$ and similarly $\theta_b$  refers to
the argument of $A_b$. In order to
differentiate the phase of $A_S$ from that of $S$ we use small and
capital letters: arg$( S )=\theta_s$, arg$(A_S)=\theta_S$. For the Higgs fields, we assume  $\theta_u=\theta_d=0$ to avoid spontaneous CP breaking (SCPV)  in the potential,  associated with a real CKM matrix \cite{Branco:2000dq},  which conflicts with experimental observations.
However, to keep our considerations as general as possible, one can also define a new phase
\begin{eqnarray}
\label{phases}\theta_{\Sigma
}=\arg{(H_u)}+\arg{(H_d)}+\arg{(S)}=\theta_u+\theta_d+\theta_s\,.
\end{eqnarray}
%The scalar state S has a VEV which breaks the
%$U(1)^{\prime}$ symmetry and generates a dynamical $\mu_{eff}=h_S
%\langle S \rangle$;
A detailed analysis of the Higgs sector
with CP violating phases is available in \cite{Demir:2003ke} and
references therein, but it is sufficient to mention that we assume $\theta_s \ne 0$, which in a more general context could be replaced by  $\theta_{\Sigma} \ne 0$.  The tree level
Higgs potential of the effective U(1)$^\prime$ model is a sum of $F$-terms, $D$-terms, and soft supersymmetry breaking terms:
\begin{equation}
\label{treepot} V_{tree}=V_D+V_F+V_{soft},
\end{equation}
where the terms $V_D,~V_F$ and $V_{soft}$ are:
\begin{eqnarray}
\label{vd}
&&V_D=\frac{g^2}{8}(|H_u|^2-|H_d|^2)^2+
   \frac{g_{2}^2}{2}(|H_u|^2|H_d|^2-|H_u \cdot H_d|^2)+
   %\nonumber \\&+&
\frac{g_{Y'}^2}{2}({\cal Q}_u|H_u|^2+{\cal Q}_d|H_d|^2+
   {\cal Q}_S|S|^2)^2, \nonumber \\
\label{vf}
&&V_F= |Y_{S}|^2\left[ |H_u\cdot H_d|^2+ |S|^2 (
|H_u|^2+|H_d|^2)\right], \nonumber \\
&&V_{soft}= m_{H_u}^{2}|H_u|^2+
  m_{H_d}^{2}|H_d|^2+
  m_S^{2}|S|^2 +(A_S Y_S S H_u\cdot H_d+h.c.),
\end{eqnarray}
where the coupling constant $g^{2}=g_2^{2}+g_Y^{2}$.  For the numerical analysis  we
take $g_Y=g_{Y^\prime}$ (the $U(1)^\prime$ coupling constant), which does not conflict with the
unification of the gauge couplings.

From the tree-level potential one can derive the minimization equations for the VEVs  $v_u,\,v_d,\, v_S$ and the phase $\theta_\Sigma(\theta_s)$. These relations yield conditions relating the VEVs to the physical Higgs masses.

The spectrum of physical Higgs
bosons consist of three neutral scalars ($h, H, H^\prime$), one CP
odd pseudoscalar ($A^0$) and a pair of charged Higgs bosons $H^\pm$ in the
CP conserving case. In total, the spectrum differs from that of
the MSSM by one extra CP-even scalar.
Notice that, the composition,  the mass and the couplings of the
lightest Higgs boson of $U(1)^{\prime}$ models can exhibit
significant differences from the MSSM, and this could be an
important source of distinguishing signatures in the forthcoming experiments. It
is important to emphasize that these models can predict naturally larger
values for $m_h$, the lightest neutral Higgs boson masses, which are more likely to agree with the boson mass seen at
the LHC. While we can safely require $m_h \ge 90$ GeV for all numerical estimates \cite{Demir:2005kg}, in principle, it is possible to
obtain larger values such as $m_h\sim 140$ GeV within some of
the  $E_6$ based models. In our evaluations, we shall impose $m_h \sim 124-126$ GeV, in agreement with the mass of the particle observed at the LHC.

\subsection{One-loop Corrections to the Higgs Potential}
\label{subsec:loopHiggs}

The tree level potential in Eq. (\ref{treepot}) is insufficient to make precise
predictions for masses and mixings, and thus we include loop corrections.  For this we
use the effective potential approach. Not all of the CP violating parameters are free parameters, and  loop
corrections induce certain relationships among them.
The one-loop corrected
potential has the form
$V=V_{tree} + \Delta V$,
where $V_{tree}$ is defined in (\ref{treepot}),  and $\Delta V$ is the one-loop Coleman-Weinberg potential \cite{Coleman:1973jx}:
\begin{equation}
\Delta V = \frac{1}{64 \pi^{2}}\left \{ \Sigma_J(-1)^{2J+1}(2J+1)
{\cal{M}}^{4}(H_u,H_d,S)\left[\ln{\frac{{\cal{M}}^{2}(H_u,H_d,S)}{\Lambda^{2}}}-
\frac{3}{2}\right]\right \},
\end{equation}
where ${\cal M}$ represent the mass matrices of all the particles in the theory.
While many particles and their superpartners could be added for the
calculation of the loop corrections, we include here the dominant
 contributions coming from the top and bottom sectors ($f=t,b$) for both the quarks and scalar quarks, so that
both contributions from small and large $\tan\beta$ values can be investigated safely. Specifically
\begin{eqnarray}
\label{colemanweinberg} \Delta V = \frac{6}{64 \pi^2}\ \sum_{f=b,t}\left\{
 \sum_{k=1,2}(m_{\widetilde{f}_{k}}^{2})^{2} \left[
\ln\left(\frac{m_{\widetilde{f}_{k}}^{2}}{\Lambda^2}\right) -
\frac{3}{2}\right] - 2(m_f^2)^{2} \left[
\ln\left(\frac{m_f^{2}}{\Lambda^2}\right) -
\frac{3}{2}\right]\right\}.
\end{eqnarray}
In this expression the masses depend explicitly on the Higgs field
components: for instance the bottom mass-squared is given by
$m_b^{2}= Y_b^{2} |H_d|^2$, and the top by $m_t^{2}= Y_t^{2} |H_u|^2$,  and the scalar quark masses-squared are
obtained by diagonalizing the mass-squared matrix,   the unitary matrix $\mathcal{S}_f$ as $\mathcal{S}_f^{\dagger}
\widetilde{M}^{2} \mathcal{S}_f =
{\mbox{diag}}(m_{\widetilde{f}_{1}}^{2},
m_{\widetilde{f}_{2}}^{2})$, with $f=t,b$.

The vacuum state is obtained by requiring the vanishing of all tadpoles
and positivity of the resulting Higgs boson masses.
 The vanishing of tadpoles for $V$ along the CP-even directions
$\phi_{H_u,H_d,S}$ and CP-odd directions $\varphi_{u,d,S}$ allows the
soft masses $m_{H_u,H_d,S}^{2}$ to be expressed in terms of the other
parameters of the potential. The tadpole terms are obtained from
\begin{align}
{\cal{T}}_{i}=\left(\frac{\partial V}{\partial \Phi_i}
\right)_{0},
\end{align}
where $_0$ means that we evaluate the derivative at the minimum of the potential, $V=V_{tree}+\Delta V$, and
$\Phi_i={\phi_u,\phi_d,\phi_S,\varphi_u,\varphi_d,\varphi_S}$.
Since all tadpole terms must vanish, enforcement of
${\cal{T}}_{1,2,3}=0$ is used to obtain $m_{H_u,H_d,S}$,
respectively, and ${\cal{T}}_{4,5,6}$ can be used for the phase of
the trilinear coupling ($A_S$), which is $\theta_S$. In fact at
the tree-level the result is $\theta_S=0$, but loop corrections
induce this quantity to be non-zero. For instance, at the tree
level,  using ${\cal{T}}_{1}$, ${\cal{T}}_{2}$ and ${\cal{T}}_{3}$ (given explicitly in the Appendices), one can express Higgs mass-squared as
\begin{eqnarray}
m_{H_u}^2&= &
\frac{A_S  Y_S  \cos(\theta_{\Sigma }+\theta_S)v_d v_S}{\sqrt{2}
v_u} -\frac{ {\cal Q}_{H_u}
\Pi+Y_S^2(v_d^2+v_S^2)}{2}+\frac{g^2(v_u^2-v_d^2)}{8}, \\
m_{H_d}^2&=&
\frac{A_S  Y_S  \cos(\theta_{\Sigma }+\theta_S)v_u v_S}{\sqrt{2}
v_d} -\frac{ {\cal Q}_{H_d}
  \Pi+Y_S^2(v_u^2+v_S^2)}{2}+\frac{g^2(v_u^2-v_d^2)}{8}, \\
m_{S}^2&= &
\frac{A_S  Y_S  \cos(\theta_{\Sigma }+\theta_S)v_d v_u}{\sqrt{2}
v_s} -\frac{ {\cal Q}_{S} \Pi+Y_S^2(v_d^2+v_u^2)}{2},
\end{eqnarray}
where
\begin{equation}
\Pi=g^2_{Y^\prime} ({\cal Q}_{H_d} v_d^2+{\cal Q}_S v_S^2+{\cal
Q}_{H_u} v_u^2).
\end{equation}

At  tree-level ${\cal{T}}_{4}$, ${\cal{T}}_{5}$ and
${\cal{T}}_{6}$ are zero,  but at one-loop level they all induce
the same non-zero result.  We collected the full form of the tadpoles
${\cal{T}}_{4}$, ${\cal{T}}_{5}$ and ${\cal{T}}_{6}$  in the
Appendices. Using the tadpoles along the CP odd directions,  the
phase of the trilinear coupling of S  ($A_S$) emerges as a
radiatively induced quantity,
\begin{align}
\label{tads} \theta_S\to -\sin^{-1}\left(\frac{3(F_b S_b A_b
Y_b^2+F_t S_t A_t Y_t^2)}{32 \pi^2
   A_S}\right)-\theta_{\Sigma },
\end{align}
where we defined $S_t=\sin (\theta_t+\theta_{\Sigma })$ and
$S_b=\sin (\theta_b+\theta_{\Sigma })$.  We define
cosine of the same quantities: $C_t=\cos (\theta_t+\theta_{\Sigma
})$ and $C_b=\cos (\theta_b+\theta_{\Sigma })$. Here
 $F_t$ and $F_b$  are loop functions:
\begin{align}
F_f=-2+\ln\left(\frac{m_{\tilde{f}_1}^2
m_{\tilde{f}_2}^2}{Q^4}\right)-\ln
\left(\frac{m_{\tilde{f}_1}^2}{m_{\tilde{f}_2}^2}\right)
\frac{\Sigma _f}{\Delta_f},
\end{align}
where $f=t,b$ refers to  top and  bottoms and we defined  $Q$ as the SUSY
breaking scale, $\Delta_f=m^2_{\tilde{f}_2}-m^2_{\tilde{f}_1}$ and
$\Sigma_f=m^2_{\tilde{f}_2}+m^2_{\tilde{f}_1}$.

%%%%%%%%%%%%%%%re-wrote up to here%%%%%%%%%

\subsection{The Higgs Mass Calculation}
\label{subsec:massHiggs}
We now turn to the Higgs mass calculation at one-loop in the
presence of CP violation in the stop and sbottom LR mixing.  The
mass-squared matrix of the Higgs scalars is
\begin{eqnarray}
{\cal{M}}^{2}_{i j}=\left(\frac{\partial^{2}}{\partial \Phi_i
\partial \Phi_j} V\right)_{0}\,.
\end{eqnarray}
In the above $\Phi_i=(\phi_i,\varphi_i)$. Two linearly
independent combinations of the pseudoscalar components
$\varphi_{u,d,S}$ are the Goldstone bosons $G_Z$ and $G_{Z'}$, which
are used to give mass to the $Z$ and $Z'$ gauge bosons, leaving one physical pseudoscalar Higgs state $A^0$, which mixes with the neutral Higgs mass states in the presence of CP violation. In the basis of scalars ${\cal{B}}=\left\{\phi_u, \phi_d,
\phi_S, A^0 \right\}$, the neutral Higgs mass-squared matrix
 ${\cal{M}}^2$ takes the following symmetric form
\begin{eqnarray}
\label{higgsmassmat}
\mathcal{M}^{2}_{H^0}=\left(
\begin{array}{cccc}
\mathcal{\mathcal{M}}^2_{11}& \mathcal{M}^2_{12}&\mathcal{M}^2_{13}&\mathcal{M}^2_{14}\\
\mathcal{M}^2_{12}& \mathcal{M}^2_{22}&\mathcal{M}^2_{23}&\mathcal{M}^2_{24}\\
\mathcal{M}^2_{13}& \mathcal{M}^2_{23}&\mathcal{M}^2_{33}&\mathcal{M}^2_{34}\\
\mathcal{M}^2_{14}&
\mathcal{M}^2_{24}&\mathcal{M}^2_{34}&\mathcal{M}^2_{44}
\end{array}\right)\, .
\end{eqnarray}
 The mass-squared matrix can be diagonalized by a $4\times 4$
orthonormal matrix ${\cal{O}}$. In doing this we follow the
convention ${\cal{O}} \mathcal{M}^{2}_{H^0}
{\cal{O}}^\dagger$=diag($m^2_{H^0_1},m^2_{H^0_2},m^2_{H^0_3},m^2_{H^0_4}$), where,
to avoid discontinuities in the eigenvalues, we
adopt the ordering: $m_{H^0_1}<m_{H^0_2}<m_{H^0_3}<m_{H^0_4}$. The elements
of ${\cal{O}}$ determine the couplings of Higgs bosons to the MSSM
fermions, scalars, and gauge bosons.

The results for the entries of the neutral Higgs (mass)$^2$ matrix are
collected in the Appendices. As an example, we show here
 one of the masses for the CP-conserving case. When CP is conserved
all $\mathcal{M}^2_{i4}$ and $\mathcal{M}^2_{4i}$  entries should
vanish, with the exception of the $\mathcal{M}^2_{44}$ term, which is actually the
pseudoscalar Higgs (mass)$^2$ term. When CP is
conserved $M^2_{A^0}$ is
\begin{align}
\mathcal{M}^2_{A^0}=&\mathcal{M}^2_{44}=\frac{\mu  \omega^2 A_S}{v_d
v_S^2 v_u}+\frac{\kappa \mu  \omega^2 \Delta_b^2 \Delta_t^2 (F_b
A_b Y_b^2+F_t A_t Y_t^2)}{v_d v_S^2 v_u},
\end{align}
where $ \omega^2=v^2 v_S^2+v_d^2 v_u^2$ and $\kappa=3/(32 \pi^2
\Delta^2_t \Delta^2_b)$.

Calculation of masses of the charged Higgs bosons is very similar to the
neutral ones and we obtain the following
mass-squared matrix
\begin{eqnarray}
\label{higgsmassmat2}
\mathcal{M}^{2}_{H^\pm}= \left(
\begin{array}{cc}
\mathcal{\mathcal{M}}^{2\,\pm}_{11}& \mathcal{M}^{2\,\pm}_{12}\\
\mathcal{M}^{2\,\pm}_{21}& \mathcal{M}^{2\,\pm}_{22}
\end{array}\right),
\end{eqnarray}
and the eigenvalue of this matrix yields, when CP is
not conserved, the expression
\begin{align}
&m^2_{H^\pm}=\frac{ \kappa \Delta_b^2 \Delta_t^2 }{3 v^2 \Sigma_b v_d v_S^2 \Sigma_t v_u}(\Sigma_t(3 Y_b^2 v_S^2(F_b \Sigma_b(\mu  A_b(C_b(v_d^4+v_u^4)+2 S_b v_d^2 v_u^2)-A_b^2 v_d v_u^3-\mu^2 v_d^3 v_u)\nonumber\\
&-\Sigma_b^2 v_d v_u^3(F_b+G_b-2)+\Delta_b^2(G_b-2) v_d v_u^3)-\Sigma_b(v_d^4+v_u^4)(8 \pi^2 v_d v_u(4 \mu^2-g_2^2 v_S^2)-\mu  \chi  v_S^2)\nonumber\\
&+6 Y_b^4 \Sigma_b v_d^3 v_S^2 v_u^3(\ln(\frac{m_b^2}{Q^2})-1))+3 \Sigma_bY_t^2 v_S^2(F_t \Sigma_t(\mu  A_t(C_t(v_d^4+v_u^4)+2 v_d^2 S_t v_u^2)-A_t^2 v_d^3 v_u-\mu^2 v_d v_u^3)\nonumber\\
&-v_d^3 \Sigma_t^2 v_u(F_t+G_t-2)+v_d^3(G_t-2) \Delta_t^2 v_u)+6
\Sigma_b v_d^3 Y_t^4 v_S^2 \Sigma_t
v_u^3(\ln(\frac{m_t^2}{Q^2})-1)).
\end{align}
where we defined the loop function
\begin{align}
 G_f=2+\ln\left(\frac{m_{\tilde{f}_1}^2}{m_{\tilde{f}_2}^2}\right)
\frac{\Sigma_f}{\Delta_f},
\end{align}
with $f=t,b$.
From this it is easy to obtain the mass of the charged Higgs in the
CP conserving case. This can be achieved by taking the limits $C_t\to 1$, $C_b\to 1$
and $S_t\to 0$, $S_b\to 0$. We present explicitly the four
entries of the charged Higgs mass-squared matrix in the Appendices.

\subsection{The Neutralino Mass Matrix in $U(1)^\prime$}
\label{subsec:neutralinos}

The presence of the CP-violating  affects
the chargino, neutralino and scalar quark mass matrices. As we are concerned here with the (tree-level) Higgs decays into neutralinos, we show the effect on the phases on the neutralino mass matrix. Note that the chargino mass matrix is unchanged from the MSSM one, though it depends on  $U(1)^{\prime}$
breaking scale through the $\mu \to \mu_{eff}$ parameter in the mass matrix. Similarly, the elements in the sfermion mass matrices are modified due to the presence of the $Z^\prime$ boson. Their explicit expressions have appeared elsewhere  \cite{Demir:2010is}.

The neutralino sector of the $U(1)^\prime$
is like the MSSM, but enlarged by a pair of higgsino and gaugino states, namely
$\tilde S$ (referred to as singlino) and ${\tilde B}^{\prime}$,
 the bare state of which we call bino-prime, while ${\tilde Z}^\prime$ (zino-prime) is  the physical mixed state. The mass matrix for the six neutralinos in the
$({\tilde B} , {\tilde W}^3 , {\tilde H}^0_d , {\tilde H}^0_u ,
{\tilde S} , {\tilde B}^{\prime})$ basis is given by a complex symmetric matrix:
%% NEUTRALINO MASS MATRIX
\bea %\footnotesize
M_{\psi^0}=\left( \begin{array}{c c c c c c c c }
M_1 &   0 & -M_Z c_{\beta} s_{W}  & M_Z s_{\beta} s_{W}  & 0  & M_K \\
0   & M_2 & M_Z c_{\beta} c_{W} & -M_Z s_{\beta} c_{W} & 0 &  0 \\
-M_Z c_{\beta} s_{W} & M_Z c_{\beta} c_{W} & 0 & -\mu_{eff} & -\mu_{\lambda} s_{\beta}& {\cal Q}_{H_d} M_{v} c_{\beta}\\
M_Z s_{\beta} s_{W} & -M_Z s_{\beta} c_{W} & -\mu_{eff} & 0 & -\mu_{\lambda} c_{\beta} & {\cal Q}_{H_u} M_{v} s_{\beta}\\
 0 & 0 & -\mu_{\lambda} s_{\beta} & -\mu_{\lambda} c_{\beta} & 0 & {\cal Q}_S M_s\\
M_K & 0 & {\cal Q}_{H_d} M_{v} c_{\beta} & {\cal Q}_{H_u} M_{v} s_{\beta} & {\cal Q}_S M_s & M_1^{\prime}\\
\end{array}\right),
 \nonumber \\
\eea
with gaugino mass parameters $M_1$ , $M_2$ , $M_1^{\prime}$
and
 $M_K$ \cite{Choi:2006fz} for $\tilde B$ ,  $\tilde W^3$ ,  ${\tilde B}^{\prime}$ and
  $\tilde B - {\tilde B}^{\prime}$ mixing respectively,  $\tan \beta=v_u/v_d$, and $\theta_W$ denotes the electroweak mixing angle. After electroweak   breaking there
  are two additional mixing parameters:
\bea M_{v} = g_{Y^{\prime}} v ~~~~~ {\rm and} ~~~~~ M_s = g_{Y^{\prime}}
v_S.
 \eea
 Moreover, the doublet-doublet higgsino  and
doublet-singlet higgsino mixing mass mixings are generated to be
\bea
\mu_{eff} = Y_S \frac{v_S}{\sqrt{2}}e^{i \theta_s} ~~~~~~~,~~~~~~~ \mu_{\lambda}
= Y_S \frac{v}{\sqrt{2}}~~,
 \eea
 where $v=\sqrt{v_u^2 + v_d^2}$. The neutralinos mass eigenstates are Majorana spinors, and they can be obtained  by diagonalization
 \begin{equation}
 \chi^0_i= {\cal N}_{ij} \psi_j~~,   \qquad {\tilde \chi}_0 = (\chi_0, {\bar \chi}^0_i )^T,
 \end{equation}
 The
neutralino mass matrix is diagonalized by the same unitary matrix
\bea
 {\cal N}^{\dagger} M_{\chi^0} {\cal N}={\rm  diag}({\tilde
m}_{\chi_1^0} ,..., {\tilde m}_{\chi_6^0} ).
 \eea
 The additional
neutralino mass eigenstates due to new higgsino and gaugino
 fields encode the effects of $U(1)^{\prime}$ models, wherever neutralinos play a
 role such as in magnetic and electric dipole moments, kaon mixing, or in Higgs decays.

\section{Constraints and Implications for the CP violating Higgs Sector}
\label{sec:constraints}

\subsection{Electric Dipole Moments}
\label{subsec:EDM}
The experimental bounds on the electric dipole moments of the neutron $d_n<6.3 \times 10^{-26}e$ cm and the electron $d_e<1.8 \times 10^{-27}e$ cm \cite{Regan:2002ta,particledata},  are some of the most tightly bound measurements in physics.  The electric dipole moment (EDM) of a spin-$\frac{1}{2}$ particle is defined from
 the effective Lagrangian \cite{Ibrahim:1997gj}
\beq
{\cal L}_I=-\frac{i}{2} d_f \bar{\psi} \sigma_{\mu\nu} \gamma_5 \psi F^{\mu\nu},
\eeq
and it is induced at the  loop level if the theory contains a source of CP violation
 at the tree level. Unlike the SM,  where the EDMs are generated through the phase of the CKM matrix at higher loop level and are thus small, in MSSM, where they are generated at one-loop level, the  electric dipole moments  are very important,  and  they provide important restrictions on the parameter space of the model. In $U(1)^\prime$ supersymmetric models, they acquire contributions from gluinos (for neutron EDM) and chargino and neutralino  (for both neutron and electron EDMs), and the  contributions are generated by $\mu_{eff}=\mu e^{i \theta_s}$, with an additional  contribution generated by the ${\tilde Z}^\prime$ neutralino. The EDM was analyzed in \cite{Demir:2003ke} in the limit in which the sfermions are much heavier than the ${\tilde Z}^\prime$.

The neutralino contributions to EDMs tends to be overall subdominant. To suppress the EDMs  we can proceed as in the MSSM \cite{Ibrahim:1997gj}: we can require that the trilinear stop coupling be mostly diagonal $A_t^{i=j} \gg A_t^{i\ne j}$ (that would suppress the sfermion contribution);  we can assume cancellation between different SUSY contributions (in particular destructive interference between  gluinos and charginos);  or we can require that the first and second generation sfermion masses be in  the TeV region. Alternatively, one can assume generically small CP-violating phases, a path we do not wish to follow here, not just based on naturalness, but because  we wish to investigate the effects of the phases on Higgs phenomenology. In the case where $g_{Y^\prime} = g_{Y}$, the case we consider here, the constraints on $U(1)^\prime$ parameters are similar to those on the MSSM. The parameter space we choose for our benchmark points insures that the contributions to the EDMs are sufficiently small.

\subsection{CP violation in $K^0-\bar{K}^0$ mixing}
 \label{subsec:epsK}

The physical phases of the Higgs singlet and in the scalar
fermion, chargino and neutralino mass matrices could alter the  the value of the measure of the CP violation in $K^0-\bar{K}^0$ mixing, measured to be $\varepsilon_K=(2.271
\pm 0.017) \times 10^{-3}$~\cite{particledata}.

The contributions to the indirect CP violation
parameter of the kaon sector,  defined as
\begin{equation} \label{eK:eq:def}
\varepsilon_K \simeq \frac{e^{i\pi/4}}{\sqrt{2}} \frac{\operatorname{Im}
\mathcal{M}_{12}}{\Delta m_K} \; ,
\end{equation}
with $\Delta m_K$ the long- and short-lived kaon mass
difference, and $\mathcal{M}_{12}$ the off-diagonal element of the neutral
kaon mass matrix, is related to the effective Hamiltonian that governs
$\Delta S=2$ transitions as
\begin{equation}
 \label{eK:M12:def}
\mathcal{M}_{12} = \frac{\vev{K^0 |\mathcal{H}_{\text{eff}}^{\Delta S=2}
|\bar{K}^0}}{2m_K} \; , \quad \text{with} \qquad
\mathcal{H}_{\text{eff}}^{\Delta S=2} = \sum_i c_i {\mathcal O}_i \; .
\end{equation}
Here $c_i$ are the Wilson coefficients and ${\mathcal O}_i$ the
corresponding four-fermion operators. In the presence of SUSY contributions, the Wilson
coefficients can be decomposed as a sum
$$c_i= c_i^W + c_i^{H^\pm}+
c_i^{\tilde{\chi}^\pm} + c_i^{\tilde{g}} + c_i^{\tilde{\chi}^0},$$
where the first contribution is the SM one, the second is the charged Higgs, and the rest are supersymmetric contributions. In $U(1)^\prime$ models, the dominant supersymmetric contributions come from the chargino mediated box diagrams, and the $\Delta S=2$
transition is largely dominated by the $(V-A)$ operator
$\mathcal{O}_1= {\bar d} \gamma^\mu P_L s {\bar d} \gamma_\mu P_L s$, similar to the MSSM, and the chargino contribution is larger than the charged Higgs contributions.  The contribution in terms of the bare chargino states is approximately \cite{Branco:2000dq}:
%Working in the weak basis for the
%$\tilde{W}-\tilde{H}$, rather than in the physical chargino basis, and
%using the mass insertion approximation for the internal squarks, in the limit of
%degenerate masses for the left-handed up-squarks, $\rm{Im}\,
%\mathcal{M}_{12}$ is given by
%
%
\begin{eqnarray}
 \label{eK:M12:equation}
\rm{Im}\, \mathcal{M}_{12} & \approx & \frac{2 G_F^2 f^2_K m_K M_W^4
}{3\pi^2\vev{m_{\tilde{q}}}^8} (V_{td}^*V_{ts}) m_t^2 \left|  \;
m_{\tilde{W}^\pm} - \cot\beta \; m_{\tilde{H}^\pm} \right| \nn \\
& & \times \left\{ \Delta A_t \sin \theta_s \;
{(m^2_{\tilde{q}})}_{12} \; I(r_{\tilde{W}^\pm}, r_{\tilde{H}^\pm}, r_{\tilde{t}_L},
r_{\tilde{t}_R}) \right\} \; ,
\end{eqnarray}
where  $f_K$ is the kaon decay constant and $m_K$ the kaon
mass; $V_{ij}$ are the $V_{CKM}$ elements, $\vev{m_{\tilde{q}}}$ is the average squark mass,
taken equal to $M_{\rm SUSY}$; $m_{\tilde{W}^\pm} = M_2$ is the wino mass,
and $m_{\tilde{H}^\pm} = \mu$ is the higgsino mass, and $r_i=m_i^2/\vev{m_{\tilde{q}}}^2$.
 The non-universality in the $LL$ soft breaking masses is parametrized
by ${(m^2_{\tilde{q}})}_{12}$, and the non-universality in the soft trilinear terms is
parametrized by $\Delta A_t\equiv A_t^{13}-A_t^{23}$. Finally, $I$ is the loop function which can be reduced to elementary functions in the limit of degenerate squark masses.
%; for the special case of $M_{\tilde W} \simeq M_{\tilde H}$ and approximately degenerate squarks, the function becomes
%$$I(x,x,1,1)=\frac12 \left[ \frac{10+19x+x^2}{3(1-x)^5}+ \frac{(1+6x+3x^2)}{(1-x)^6}\ln x \right].$$
%%

Scanning the parameter space of the model, we checked that one can
find  parameter sets that satisfy the minimization conditions of the Higgs
potential, have an associated Higgs boson spectrum compatible with the LHC boson,
and still succeed in obeying the bound for the observed value of $\varepsilon_K$. From
Eq.~(\ref{eK:M12:equation}), it appears that $\eK$ depends on $1/M_{\rm
SUSY}^8$. To satisfy the experimental value of $\eK$,
 values of $M_{\rm SUSY} \ge 1$ TeV would have to be assumed, or $\Delta A_t\equiv A_t^{13}-A_t^{23} \ll 1$, in agreement with our EDM considerations.  Too small values of
$M_{\rm SUSY}$ might generate a light Higgs boson spectrum already excluded by
LEP and LHC, and for $M_{\rm SUSY} \sim 1$ TeV, which is consistent with our squark and slepton masses, the supersymmetric contributions to $\eK$ are consistent with the  experimental constraints.

\section{Numerical Analysis}
\label{sec:numerical}

As mentioned in the introduction, gauge extensions of the SM by one or several non-anomalous $U(1)^\prime$ gauge groups can arise naturally from a string-inspired $E_6SSM$ model \cite{Hall:2011au,Hall:2011zq}.  In $E_6SSM$ models the matter sector includes a {\bf 27}-representation for each family of quarks and leptons (including right-handed neutrinos), Higgs representations (doublets $H_u$ and $H_d$ and singlet $S$), and three-families of extra down-like color triplets. Anomaly cancellation occurs generation by generation, and gauge coupling unification requires another pair of Higgs-like multiplets. Breaking of $E_6$ yields $SU(3) \times SU(2) \times U(1)_Y \times U(1)^\prime$ as a low energy group. Anomaly-free $U(1)^\prime$ groups are thus generated this way, directly, or as a specific linear combination.  We first  define the models that shall be investigated in our numerical analysis. They all emerge from breaking of higher groups \cite{Langacker:2008yv}.
For instance, the anomaly free groups $U(1)_\psi$ \cite{Langacker:1998tc} and $U(1)_\chi$ \cite{Hewett:1988xc} are defined by:
$$E_6 \to SO(10) \times U(1)_\psi, \quad SO(10) \to SU(5) \times U(1)_\chi. $$
In general a $U(1)^\prime \equiv U(1)_{E_6}$ group is defined as $U(1)_{E_6}= \cos \theta_{E_6}U(1)_\chi+\sin \theta_{E_6} U(1)_\psi$, and we distinguish among the different scenario by the values of $\theta_{E_6}$:
\begin{itemize}
\item $\theta_\eta=\pi-\arctan\sqrt{\frac53}$ for $U(1)_\eta$ wich occurs in Calabi-Yau compactification of heterotic strings \cite{Witten:1985xc};
\item $\theta_S=\arctan\sqrt{15}/9$ for the secluded $U(1)_S$, where the tension between the electroweak scale and developing a large enough $Z^\prime$ mass is resolved by the inclusion of additional singlets  \cite{Erler:2002pr};
\item $\theta_I=\arctan\sqrt{\frac35}$ for the inert $U(1)_I$, which has a charge orthogonal to ${\cal Q}_\eta$ \cite{Robinett:1982tq};
\item $\theta_N= \arctan \sqrt{15}$ for $U(1)_N$, where $\nu^c$ has zero charge, allowing for large Majorana masses \cite{King:2005jy,Barger:2003zh};  and
\item $\displaystyle \theta_\psi=\frac{\pi}{2}$ for $U(1)_\psi$,  defined above from the breaking of $E_6$ \cite{Langacker:1998tc}.
\end{itemize}
In the Table \ref{tab:E6decomp}   below,  we list the charges for the fundamental representations of $E_6$  in the $U(1)^\prime$ models which we use for numerical investigations of Higgs boson properties.
\begin{table*}[ht]
\begin{center}
\begin{tabular}{|c| c| c| c| c|c|c|}
\hline $SO(10)$ representations & $SU(5)$ representations& $2 \sqrt{15} {\cal Q}_{\eta}$ &  $2 \sqrt{15} {\cal Q}_S$& $2{\cal Q}_I$ &$2 \sqrt{10} {\cal Q}_N$ &  $2 \sqrt{6}
{\cal Q}_{\psi}$    \\
\hline
{\bf 16}   &   ${\bf 10}~ (u,d,{u^c}, {e^+} )$ & $-2$ & $-1/2$  & $0$  &1& $1$\\
 $(u,d, \nu, e^-, u^c,d^c, \nu^c, e^+)$           &   ${{\bf 5}^\ast}~ ( d^c, \nu ,e^-)$  & 1  & 4 & $-1$  & 2  & 1  \\
            &   $\nu^c$             & $-5$ & -$5$    &1  & 0 & 1      \\
\hline
       {\bf 10}   &   ${\bf 5}~(H_u)$    & 4  & $1$ & 0  & $-2$    & $-2$  \\
 $(H_u, H_d)$           &   ${{\bf 5}^\ast} ~( H_d)$ & $1$ &$-7/2$ & 1& $-3 $ & $-2$\\
\hline
       {\bf 1} (S)   &   ${\bf 1}~ (S)$                  &  $-5$ & $5/2$ & $-1$ & 5 & $4$\\
\hline
\end{tabular}
\end{center}
\caption{\label{tab:E6decomp} Values of $U(1)_{\eta}$,  $U(1)_S$, $U(1)_I$,
 $U(1)_N$ and  $U(1)_{\psi}$ charges for the ${\bf 27}$ fundamental representation of $E_6$  decomposition
 under $SO(10)$ and $SU(5)$ representations.
The
charge for each model is defined as ${\cal Q}=\cos\theta_{E_6}
{\cal Q}_\chi+\sin\theta_{E_6} {\cal Q}_{\psi}$.}
\end{table*}

In what follows, we investigate the consequences of each of the anomaly-free groups on the Higgs production and decay at the LHC. In Table \ref{tab: cp inputs} we list the relevant benchmark parameters for each of the choices, for both the CP-violating  (CP-conserving) Higgs sectors{\footnote {By CP violating scenario, we mean the specific case where $\theta_s$ is given by the values in Table \ref{tab: cp inputs}.}}. In addition to the phase $\theta_s$ (which defines the CP violating scenario of each model), the values of $\tan \beta$ and of $\mu$, we give the $U(1)_Y$ and $SU(2)_L$ gaugino masses $M_1$ and $M_2$, the left and right handed squark soft mass parameters $M_{Q_i}$ and $M_{U_i}$ (all taken to be 1 TeV, including the masses in the down scalar sector, not explicitly shown), the trilinear couplings in the top and bottom scalar quark sectors, $A_t$ and $A_b$, and the ratios $\displaystyle R_{Y'}=\frac{M_1^\prime}{M_1}$ and $\displaystyle R_{YY'}=\frac{M_K}{M_1}$, as defined in  \cite{Demir:2010is}. The constraints on the mass parameters, constraining the choice of benchmark values, are:
\begin{itemize}
\item Requiring the lightest  Higgs mass to be very close to 126 GeV, in agreement with the ATLAS and CMS results;
\item Requiring the next lightest neutral Higgs boson to have mass $m_{H^0_2} >600$ GeV (as it has not been observed at LHC);
\item Requiring the lightest neutralino mass to be consistent with collider limits on Z boson decays, but also to allow for the possibility of the neutral Higgs boson to decay into a neutralino pair;
\item Choosing the lightest neutralino to be the LSP and requiring that the relic density constraint be satisfied;
\item Choosing the $Z^\prime$ boson mass to be consistent with present limits \cite{particledata};
\item Choosing scalar masses and trilinear couplings which satisfy constraints from EDMs and CP violation in the kaon sector, as described in the previous section (\ref{sec:constraints}).
\end{itemize}
As we would like to allow the Higgs boson to be kinematically allowed to decay into two neutralinos, we impose the LEP constraint on the $Z$ boson width \cite{LEP} $\Gamma (Z \to {\tilde \chi}_1^0 {\tilde \chi}_1^0) < 3$ MeV. This constraint allows for a weakening of the Particle Data bound \cite{particledata}, especially as we do not impose the supersymmetric grand unified theory relationship $M_1=(5/3)\tan^2 \theta_W M_2$, and allow $M_1$ and $M_2$ to be free parameters, as given in Table \ref{tab: cp inputs}. Note that in particular, the bino mass is chosen to be light to allow Higgs decays into neutralinos, while value for $M_2$ insures that the chargino mass will be $m_{{\tilde \chi}^\pm_1} >m_{H^0_1}/2$. The scalar fermions are heavy to satisfy bounds from the EDMs and $\varepsilon_K$. We choose the value of $\theta_s$ for each model to maximize the invisible decay width for the lightest Higgs boson, while satisfying the other constraints\footnote{Our benchmarks are different from those NMSSM \cite{Djouadi:2008uw}, where CP conservation was assumed, and where the dominant decay mode of the lightest CP-even Higgs is into the pseudoscalar Higgs boson pairs.}.

%%%%%%%%%%%%%%%%%%%%%%%%%%%%%%%%%%%%%%%%%%%%%%%%%%%%%%%%%%%%%%%%%%%%%%%%%%%%%%%%%%%%%%%%%%
\setlength{\voffset}{-0.5in}
\begin{table}[htbp]
 \begin{center}
\setlength{\extrarowheight}{-5.8pt} \small
\begin{tabular*}{0.95\textwidth}{@{\extracolsep{\fill}} cccccc}
\hline\hline
 $\rm Parameters$ & $U(1)_{\eta}$  &$U(1)_{S}$ &$U(1)_{I}$&$U(1)_{N}$&$ U(1)_{\psi}$
 \\ \cline{1-5}\cline{1-6}
 %$\bf{\theta_6}$&\bf{127.8}&\bf{23.4}&\bf{37.8}&\bf{75.6}&\bf{90.0}\\
 $\theta_s$&42(0)&75(0)&60(0)&55(0)&33(0)\\
 $\tan\beta$&1.8(1.7)&1.46(1.42)&1.3(2.5)&1.5(1.8)&1.75(1.75)\\
 %$h_s$&0.1(0.1)&0.1(0.1)&0.1(0.1)&0.1(0.1)&0.1(0.1)\\
 %$A_s$&600(600)&600(600)&600(600)&600(600)&600(600)\\
 $\mu(|\mu_{eff}|)$&360(360)&715(730)&465(461)&292(295)&285(290)\\
 $M_1$&48(50)&56(59)&57(50)&49(50)&49(51)\\
 $M_2$&125(130)&115(120)&135(120)&130(170)&140(160)\\
 %$M_3$&1000(1000)&1000(1000)&1000(1000)&1000(1000)&1000(1000)\\
 $M_{Q_1}$&1000(1000)&1250(850)&750(600)&2000(300)&1000(1000)\\
 $M_{Q_2}$&1000(1000)&1250(850)&750(600)&2000(300)&1000(1000)\\
 $M_{Q_3}$&1000(1000)&1250(850)&750(600)&2000(300)&1000(1000)\\
 $M_{U_1}$&1000(1000)&1250(850)&750(600)&2000(300)&1000(1000)\\
 $M_{U_2}$&1000(1000)&1250(850)&750(600)&2000(300)&1000(1000)\\
 $M_{U_3}$&1000(1000)&1250(850)&750(600)&2000(300)&1000(1000)\\
 %$M_{Q_1}$&1000(1000)&1000(1000)&1000(1000)&1000(1000)&1000(1000)\\
 %$M_{Q_2}$&1000(1000)&1000(1000)&1000(1000)&1000(1000)&1000(1000)\\
 %$M_{Q_3}$&1000(1000)&1000(1000)&1000(1000)&1000(1000)&1000(1000)\\
 %$M_{U_1}$&1000(1000)&1000(1000)&1000(1000)&1000(1000)&1000(1000)\\
 %$M_{U_2}$&1000(1000)&1000(1000)&1000(1000)&1000(1000)&1000(1000)\\
 %$M_{U_3}$&1000(1000)&1000(1000)&1000(1000)&1000(1000)&1000(1000)\\
 %$M_{D_1}$&1000(1000)&1000(1000)&1000(1000)&1000(1000)&1000(1000)\\
 %$M_{D_2}$&1000(1000)&1000(1000)&1000(1000)&1000(1000)&1000(1000)\\
 %$M_{D_3}$&1000(1000)&1000(1000)&1000(1000)&1000(1000)&1000(1000)\\
 $|A_t|$&1850(2000)&2200(2500)&2500(1500)&2250(2000)&2000(2000)\\
 $|A_b|$&2000(2000)&2500(2500)&2500(1500)&2500(2000)&2000(2000)\\
 $R_{Y^\prime}$&1(1)&2.5(0.1)&0.1(6.6)&1(1)&5(5)\\
 $R_{Y Y^\prime}$&1(2.2)&0.1(0.1)&2(6.6)&6(2.7)&0.1(5)\\
 %$\sin\chi$&0.001(0.001)&0.001(0.001)&0.001(0.001)&0.001(0.001)&0.001(0.001)\\
 \hline\hline
\end{tabular*}
\caption{\label{tab: cp inputs}\sl\small The benchmark points (in GeV) for
the  CP-violating (CP-conserving) $U(1)_{\eta}$, $U(1)_S$, $U(1)_I$, $U(1)_N$ and
$U(1)_{\psi}$
   versions of  $U(1)^{\prime}$
models.}
\end{center}
 \end{table}
%%%%%%%%%%%%%%%%%%%%%%%%%%%%%%%%%%%%%%%%%%%%%%%%%%%%%%%%%%%%%%%%%%%%%%%%%%%%%%%

\begin{figure}[htb]
%\vskip -0.3in
\begin{center}$
 \begin{array}{cc}
         \hspace*{-0.6cm}
      % \hspace*{-1.7cm}
  \includegraphics[width=2.7in,height=2.5in]{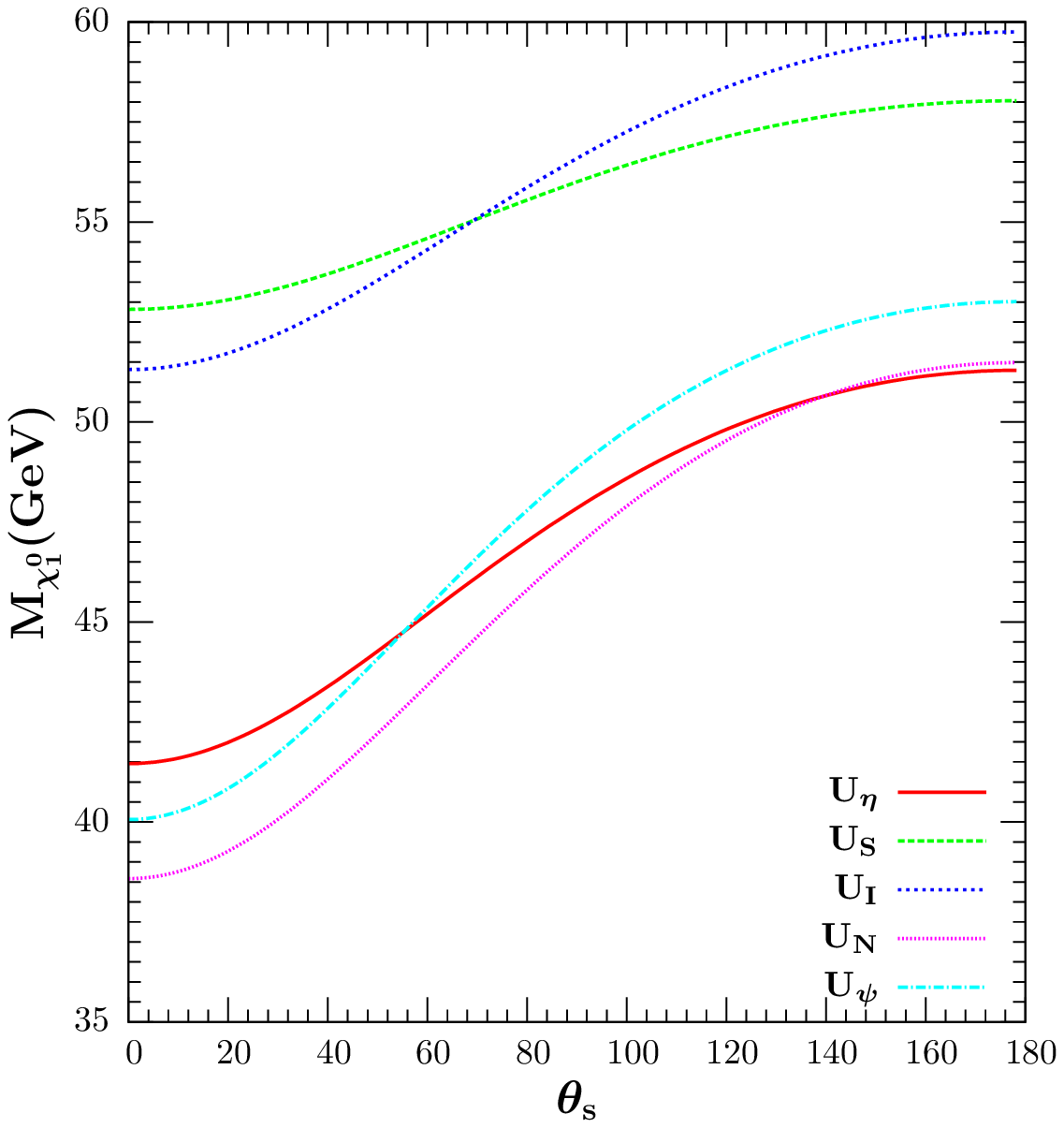}
&\hspace*{1.0cm}
       \includegraphics[width=2.8in,height=2.5in]{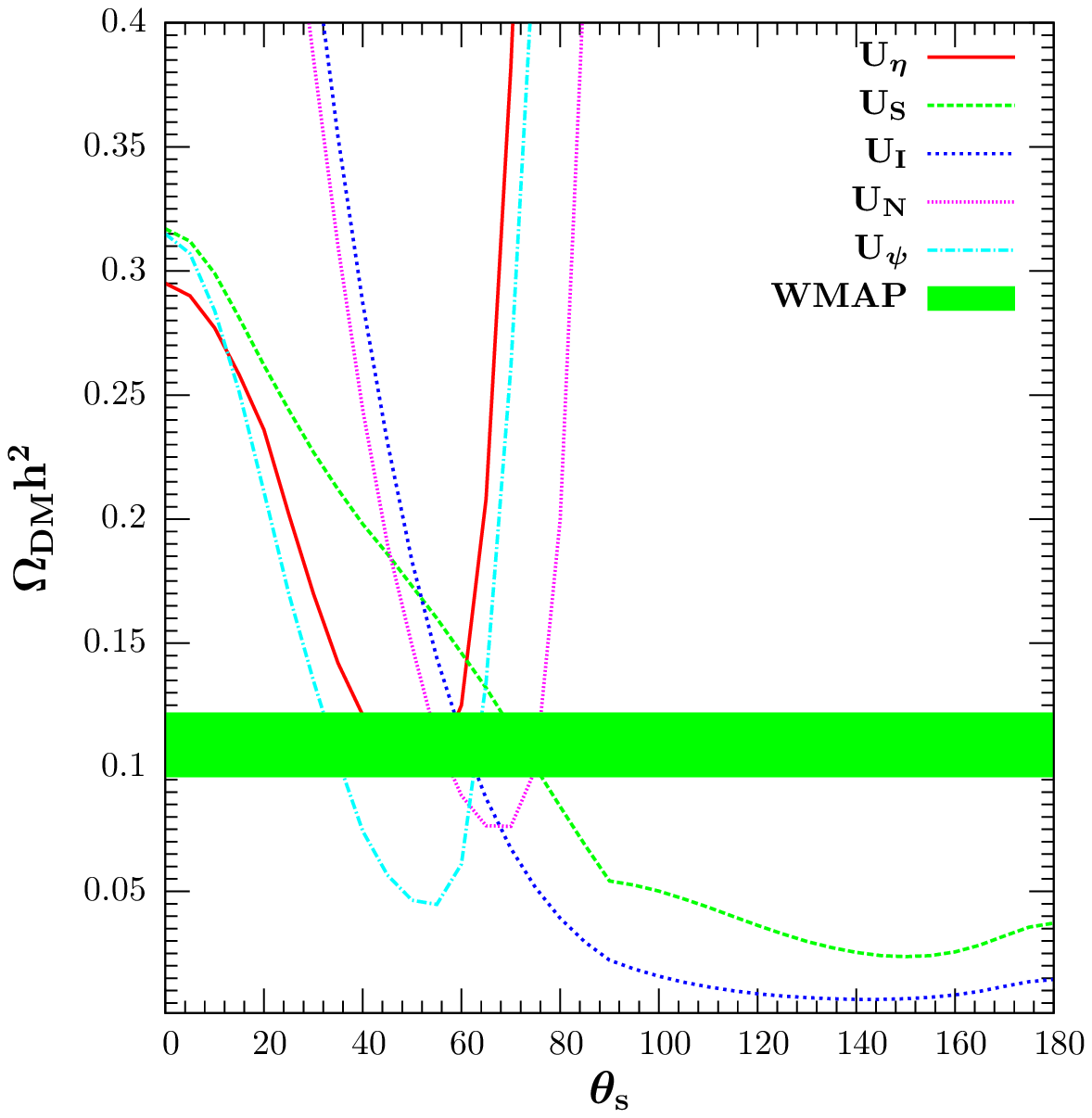}
       \end{array}$
\end{center}
\vskip -0.3in
     \caption{Mass of the lightest neutralino and the relic density  as  functions of  \sl\small $\theta_s$ (the phase of
the new singlet S) for the CP-violating versions of $U(1)_{\eta}$, $U(1)_S$, $U(1)_I$, $U(1)_N$ and
$U(1)_{\psi}$
models. The green band indicates the experimentally allowed region. }
\label{fig:mchi1theta}
\end{figure}

Based on the input parameters, we calculate the spectrum of the physical masses of the extra particles in the model, which are used in our numerical evaluations. These values are given in
 Table \ref{tab: cp masses}. We also included in this table  the relic density of the
dark matter for all scenarios. Throughout our considerations the lightest neutralino ${\tilde \chi}_1^0$ is the lightest supersymmetric particle (LSP) and thus subject to cosmological constraints.  The relic calculation is straightforward
using the {\tt Micromegas} package \cite{Belanger:2008sj}, once we
include the $U(1)^\prime$ model files from
 {\tt CalcHEP} \cite{calchep}. All the numbers are within the $1\sigma$ range of the WMAP
result \cite{wmap}  from the Sloan
Digital Sky Survey \cite{Spergel:2006hy}
 \begin{eqnarray}
\Omega_{DM} h^2 = 0.111^{+0.011}_{-0.015}\,.
\end{eqnarray}
The relic density of the dark matter $\Omega_{\rm DM} h^2$ is
very sensitive to the free parameter $R_{Y^\prime}$ listed in
Table~\ref{tab: cp inputs}.
%%%%%%%%%%%%%%%%%%%%%%%%%%%%%%%%%%%%%%%%%%%%
\setlength{\voffset}{-0.5in}
\begin{table}[htbp]
 \begin{center}
\setlength{\extrarowheight}{-5.8pt} \small
\begin{tabular*}{0.95\textwidth}{@{\extracolsep{\fill}} cccccc}
\hline\hline
 $\rm Masses$ & $U(1)_{\eta}$  &$U(1)_{S}$ &$U(1)_{I}$&$U(1)_{N}$&$ U(1)_{\psi}$
 \\ \cline{1-5}\cline{1-6}
 $ m_{Z^\prime}$&1510(1510)&1507(1539)&1513(1500)&1502(1517)&1513(1540)\\
 $m_{\tilde\chi^0_1}$&43(43)&55(55)&54(44)&43(42)&42(42)\\
 $m_{\tilde\chi^0_2}$&108(109)&112(110)&125(107)&111(137)&114(128)\\
 $m_{\tilde\chi^0_3}$&361(361)&715(730)&464(463)&292(297)&286(292)\\
 $m_{\tilde\chi^0_4}$&386(388)&726(742)&485(479)&326(336)&322(331)\\
 $m_{\tilde\chi^0_5}$&1487(1489)&1440(1536)&1514(1378)&1505(1498)&1396(1438)\\
 $m_{\tilde\chi^0_6}$&1535(1540)&1580(1543)&1521(1711)&1556(1549)&1641(1694)\\
 $m_{\tilde\chi^\pm_1}$&107(107)&111(110)&124(106)&108(134)&111(125)\\
 $m_{\tilde\chi^\pm_2}$&382(384)&724(740)&481(477)&321(332)&318(326)\\
 $m_{H_1^0}$&125.0(125.0)&125.6(125.0)&125.8(126.0)&125.6(126.0)&125.4(125.0)\\
 $m_{H_2^0}$&743(747)&969(1027)&788(930)&642(688)&665(679)\\
 $m_{H_3^0}$&750(754)&977(1033)&798(933)&652(695)&673(687)\\
 $m_{H_4^0}$&1510(1510)&1508(1539)&1513(1500)&1502(1517)&1513(1540)\\
 $m_{H^\pm}$&572(543)&717(711)&507(802)&418(504)&486(486)\\
 $m_{\tilde e_L}$&1341(1341)&1837(1616)&1306(1219)&700(742)&1134(1139)\\
 $m_{\tilde e_R}$&1054(1054)&1154(695)&748(598)&513(564)&1133(1137)\\
 $m_{\tilde \mu_L}$&1341(1341)&1837(1616)&1306(1219)&700(742)&1134(1139)\\
 $m_{\tilde \mu_R}$&1054(1054)&1154(695)&748(598)&513(564)&1133(1137)\\
 $m_{\tilde \tau_1}$&1054(1054)&1154(695)&748(598)&513(564)&1133(1137)\\
 $m_{\tilde \tau_2}$&1342(1341)&1837(1616)&1306(1219)&700(742)&1135(1139)\\
 $m_{\tilde\nu_e}$&1340(1340)&1836(1615)&1306(1217)&699(739)&1133(1137)\\
 $m_{\tilde\nu_{\mu}}$&1340(1340)&1836(1615)&1306(1217)&699(739)&1133(1137)\\
 $m_{\tilde\nu_{\tau}}$&1340(1340)&1836(1615)&1306(1217)&699(739)&1133(1137)\\
$m_{\tilde u_{L}}$&1054(1054)&879(874)&999(998)&1106(1108)&1133(1137)\\
$m_{\tilde u_{R}}$&1055(1055)&882(877)&1001(1000)&1107(1109)&1134(1138)\\
$m_{\tilde d_{L}}$&1056(1055)&880(875)&1000(1001)&1107(1109)&1134(1139)\\
$m_{\tilde d_{R}}$&1340(1340)&1675(1698)&1463(1457)&1203(1207)&1133(1138)\\
$m_{\tilde c_{L}}$&1054(1054)&879(874)&999(998)&1106(1108)&1133(1137)\\
$m_{\tilde c_{R}}$&1055(1055)&882(877)&1001(1000)&1107(1109)&1134(1138)\\
$m_{\tilde s_{L}}$&1056(1055)&880(875)&1000(1001)&1107(1109)&1134(1139)\\
$m_{\tilde s_{R}}$&1340(1340)&1675(1698)&1463(1457)&1203(1207)&1133(1138)\\
$m_{\tilde t_{1}}$&919(911)&659(670)&788(894)&938(968)&994(1002)\\
$m_{\tilde t_{2}}$&1201(1207)&1085(1070)&1200(1122)&1277(1275)&1281(1283)\\
$m_{\tilde b_{1}}$&1056(1055)&880(875)&1000(1001)&1107(1109)&1130(1135)\\
$m_{\tilde b_{2}}$&1340(1340)&1675(1698)&1463(1457)&1203(1207)&1137(1141)\\
\hline
$\Omega_{DM}$&$0.114(0.120)$&$0.100(0.102)$&$0.113(0.120)$&$0.111(0.117)$&$0.117(0.101)$\\
\hline\hline
\end{tabular*}
\caption{\label{tab: cp masses}\sl\small The mass spectra (in GeV) and the relic density $\Omega_{\rm DM}$ values
for the CP-violating (CP-conserving) version of the scenarios
considered  given  in Table~\ref{tab: cp inputs}
for the $U(1)^{\prime}$ models.}
\end{center}
 \end{table}
%%%%%%%%%%%%%%%%%%%%%%%%%%%%%%%%%%%%%%%%%%%%%%%%%%%%%%%%%%
As the lightest neutralino plays an essential role in the decay of the lightest Higgs boson, we first show the dependence of its mass, and of the relic density with the  CP violating parameter $\theta_s$ in Fig. \ref{fig:mchi1theta}.  In all of the $U(1)^\prime$ models under study the lightest neutralino is mostly bino. The variations of its mass and of the relic density with the other CP violating phase $\theta_t$ are negligible. Note that the mass of the LSP increases smoothly with increasing $\theta_s$, while the relic density measurement (shown as a green band in the right-handed part of the plot) poses restrictions on the combined LSP mass and CP violating parameter. The values of $\theta_s$ for various models listed in Table \ref{tab: cp inputs} fall into the range of the values allowed by the relic density (within the green band). We incorporate these restrictions in our analysis of Higgs mass and decay widths.

We proceed to examine the effects of the CP violating phases on the masses, production cross sections and branching ratios of the lightest Higgs boson.
%%%%%%%%%%%%%%%%%%%%%%%%%%%%%%%%%%%%%%%%%%%%%%%%%%%%%%%%%%%%%%%%%%
\subsection{ The lightest CP-even neutral Higgs boson}
\label{subsec:H1}

The observation of the new boson at the LHC has fueled speculations of its nature (is it or not the SM Higgs boson), coupled with analyses of its mass and couplings, and their comparison with the experimental data. ATLAS \cite{ATLASupdate} and CMS \cite{CMSupdate} have reported updates on the combined strength values for main channels, including $H^0 \to b {\bar b}, \gamma \gamma, \tau^+ \tau^-, WW^* (\to \ell \nu \ell \nu) $ and  $ZZ^* (\to 4 \ell)$. While the results still have significant experimental and systematic uncertainties, these are expected to decrease with LHC operating at $\sqrt{s}=14$ TeV and increased luminosity. The precise determination of the Higgs couplings to different channels will establish whether the boson observed at the LHC is the SM Higgs boson. In our analysis, we wish to explore the possibility that Higgs boson decays in a non-SM fashion, in particular, that it can decay significantly invisibly.  A invisible decay mode is very hard to  measure directly at the colliders. However, it is not difficult to be inferred indirectly. The total decay width of a SM Higgs boson with mass of 125 GeV is approximately $\Gamma_{H^0}= 4.2$ GeV. A discrepancy between the theoretical and experimental value for the width would be an indication of additional decay channels beyond SM. Similarly, reduced decay branching ratios into known SM Higgs decay modes, in particular for $H^0  \to b {\bar b}$ and $H^0 \to  \tau^+ \tau^-$ (dominant for $m_{H^0}=126$ GeV), could also indicate that other decays are important.

\begin{figure}[htb]
%\vskip -0.3in
\begin{center}$
 \begin{array}{ccc}
         \hspace*{-1.6cm}
      \includegraphics[width=2.5in,height=2.5in]{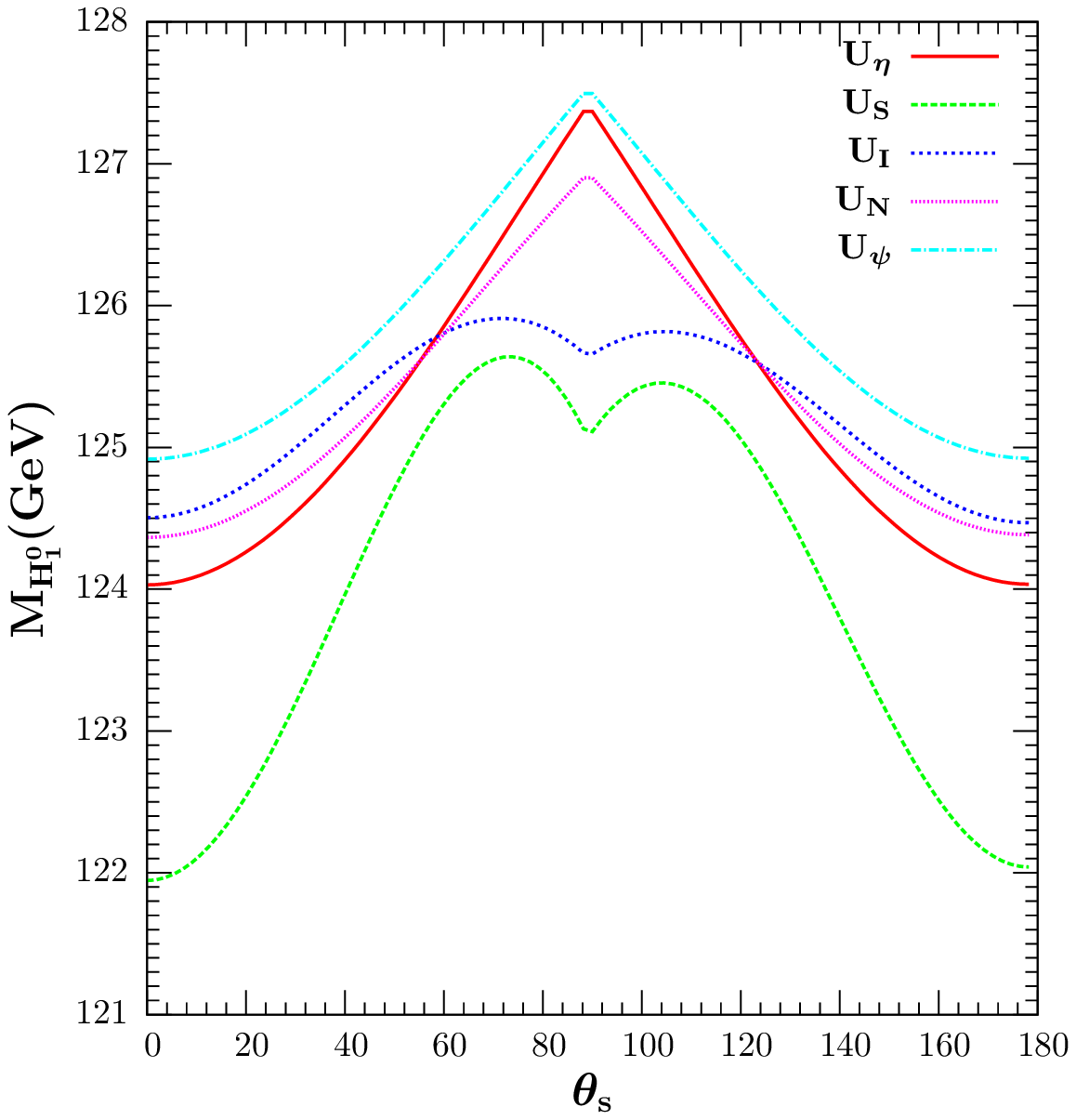}
&\hspace*{-0.4cm}
    \includegraphics[width=2.5in,height=2.5in]{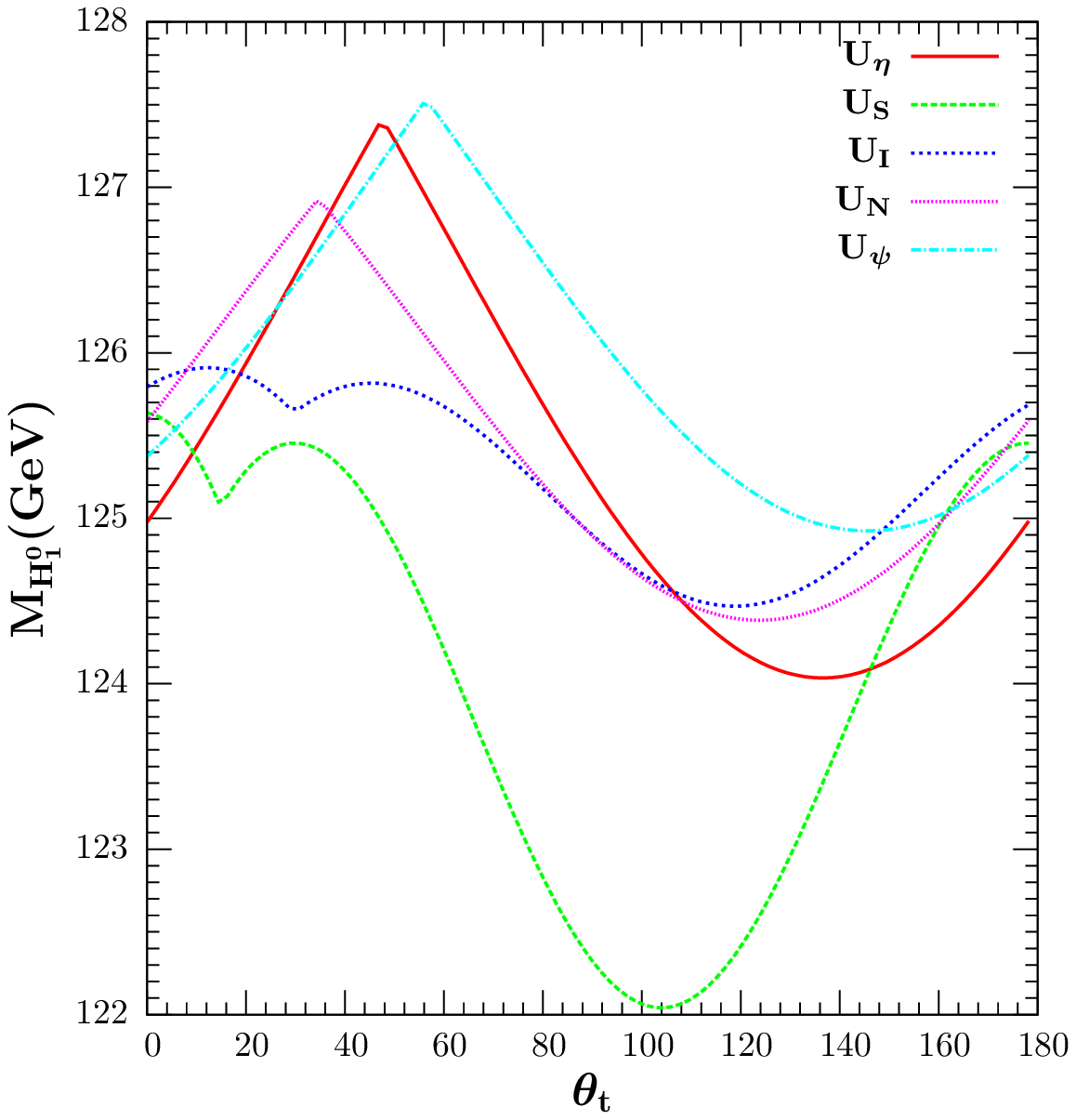}
    &\hspace*{-0.3cm}
    \includegraphics[width=2.5in,height=2.5in]{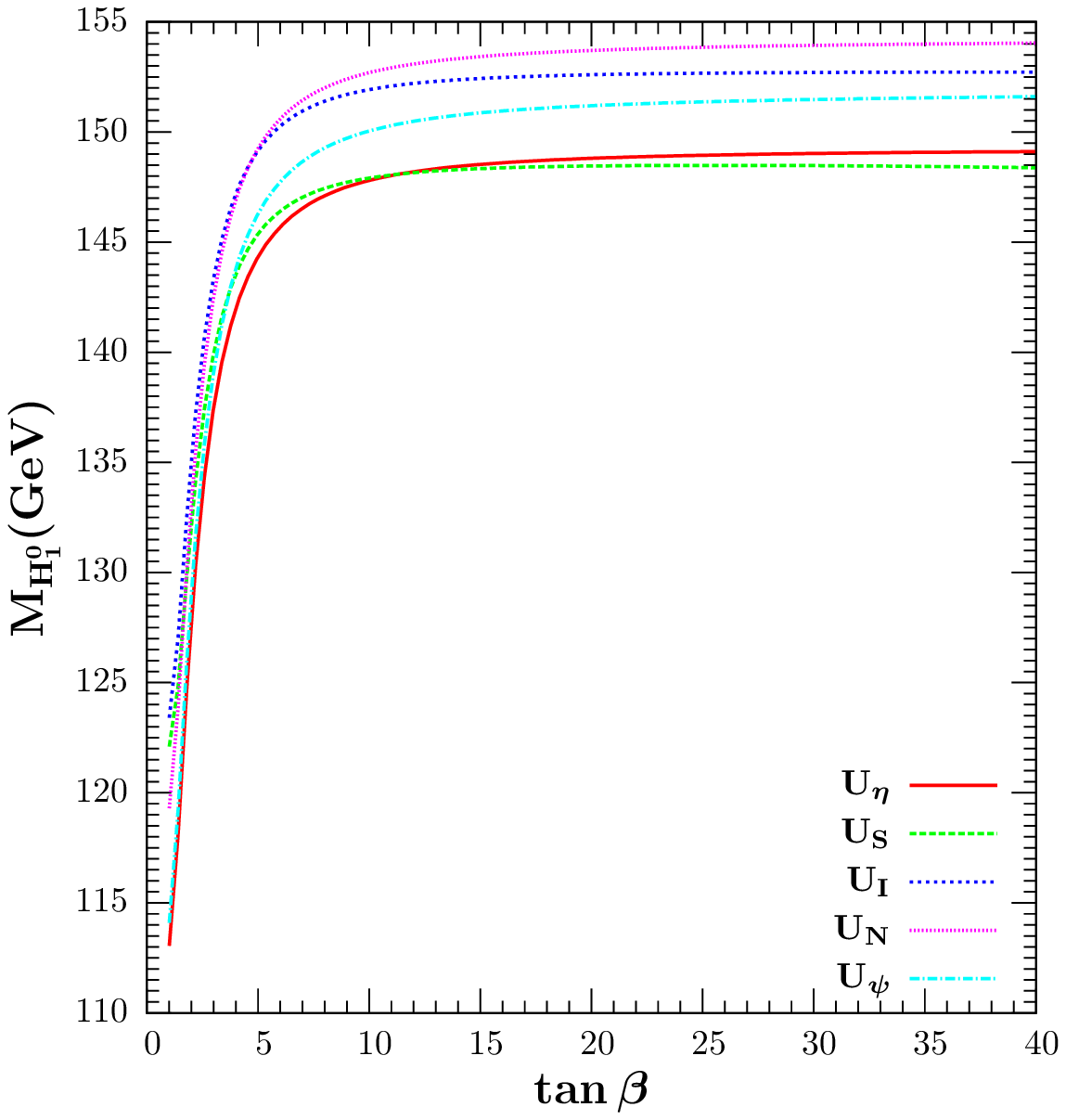}
       \end{array}$
\end{center}
\vskip -0.3in
     \caption{Mass of the lightest neutral Higgs boson as a function of  \sl\small $\theta_s$ (the phase of
the new singlet S),  $\theta_t$ (the phase of the soft coupling
$A_t$),  and $\tan \beta$ for the CP-violating versions of $U(1)_{\eta}$, $U(1)_S$, $U(1)_I$, $U(1)_N$ and
$U(1)_{\psi}$
models.}
\label{fig:MH1theta}
\end{figure}

At $\tan \beta \approx 1$ the lightest Higgs mass is determined mostly by the new $F$- and $D$-terms in the Higgs tree-level potential, and is thus sensitive to the trilinear Yukawa coupling $Y_S$ and the gauge coupling $g_{Y^\prime}(=g_Y)$ in the numerical analysis.
We first present our results for the dependence of the masses on the CP violating phases arg ($\mu_{eff})=\theta_s$ and arg ($A_t)=\theta_t$, as well as with $\tan \beta$ in Fig. \ref{fig:MH1theta}.  One can see that the mass variations with $\theta_s$ and $\theta_t$ are significant, especially in $U(1)_S$, where large regions of the parameter space for both phases, if combined with other measurements,  can be eliminated. The dependence on $\tan \beta$ from the third panel of the figure seems to indicate that only low values $\tan \beta \approx 1-2$ are allowed for all $U(1)^\prime$ models, in agreement with the values chosen in Table \ref{tab: cp inputs}.

To analyze the decay width of the lightest Higgs boson, we first calculate total production cross section  of the lightest Higgs boson ($H^0_1$) in various models  in Table \ref{tab: cp cross sec}, for $\theta_s=0$ (no CP violation) and for $\theta_s$  as in Table \ref{tab: cp inputs} (with CP violation).
We list associated Higgs-vector boson cross sections, and the total cross section for the vector boson fusion. Though subdominant production modes for Higgs bosons, these are the dominant channels for observing an invisible decay of the Higgs boson \cite{Godbole:2003it}.  Note that we do not include here the dominant production mechanism $gg \to H_1^0$, as this mode is plagued by large QCD corrections, and thus it is difficult to isolate the invisible decay of the Higgs boson, which in this production channel is expected to come from $gg \to H_1^0+$jet, and be small. As expected, the vector-boson fusion production mechanism dominates over the Higgs-vector boson associated production in all models. The numbers are fairly consistent across the models, and largely independent of CP violating phases. Thus we forgo plots of the production cross sections and expect that  any differences  would show up in the branching ratios of the lightest Higgs boson.
%%%%%%%%%%%%%%%%%%%%%%%%%%%%%%%%%%%%%%%%%%%%%%%%%%%%%%%%%%%%%%%%%%%%%%%%%%%%%%%%%%%%%%%%%%%
\setlength{\voffset}{-0.5in}
\begin{table}[htbp]
 \begin{center}
\setlength{\extrarowheight}{-5.8pt} \small
\begin{tabular*}{0.95\textwidth}{@{\extracolsep{\fill}} cccccc}
\hline\hline
 $\rm Observables$ & $U(1)_{\eta}$  &$U(1)_{S}$ &$U(1)_{I}$&$U(1)_{N}$&$ U(1)_{\psi}$
 \\ \cline{1-5}\cline{1-6}
$\rm \sigma(pp\rightarrow H^0_1Z)$&$639(642)$&$631(647)$&$628(610)$&$628(624)$&$634(642)$\\
$\rm \sigma(pp\rightarrow H^0_1W^+)$&$720(725)$&$708(725)$&$705(687)$&$708(701)$&$711(720)$\\
$\rm \sigma(pp\rightarrow H^0_1W^-)$&$445(447)$&$437(448)$&$435(424)$&$437(433)$&$439(444)$\\
%$\rm \sigma(pp\rightarrow H_1H_2)/fb$&$2(2)$&$0.4(0.2)$&$1(0.4)$&$5(4)$&$6(5)$\\
%$\rm \sigma(pp\rightarrow H_1Z^\prime)/fb$&$3.3(3.8)\times10^{-4}$&$3.4(2.0)\times10^{-4}$&$4.6(0.37)\times10^{-4}$&$1.3(1.2)\times10^{-3}$&$2.1(1.9)\times10^{-3}$\\
%$\rm \sigma(pp\rightarrow H_1H^+)/fb$&$8.6(10)\times10^{-4}$&$2.4(1.3)\times10^{-4}$&$1.9(0.06)\times10^{-3}$&$4.6(1.9)\times10^{-3}$&$2.2(1.5)\times10^{-3}$\\
%$\rm \sigma_{TOT}(pp\rightarrow H^0_1X)$&34900(37880)&67200(33980)&37100(35790)&42190(39810)&44060(31680)\\
$\rm \sigma(pp\rightarrow H^0_1jj (VBF))$&4983(4930)&4848(4920)&4861(4840)&4874(4850)&4873(4893)\\
\hline\hline
\end{tabular*}
\caption{\sl\small Total cross sections of associated production
channel ($H^0_1X$) and  vector boson fusion production channel
($H^0_1jj$) (in $fb$)
for the CP-violating (CP-conserving) versions of $U(1)_{\eta}$, $U(1)_S$, $U(1)_I$, $U(1)_N$ and
$U(1)_{\psi}$ models
   considered in the paper. } \label{tab: cp cross sec}
\end{center}
 \end{table}
%%%%%%%%%%%%%%%%%%%%%%%%%%%%%%%%%%%%%%%%%%%%%%%%%%%%%%%%%%%%%%%%%%%%%%%%%%%%%%%

\begin{figure}[htb]
%\vskip -0.3in
\begin{center}$
 \begin{array}{ccc}
         \hspace*{-1.6cm}
      \includegraphics[width=2.5in,height=2.5in]{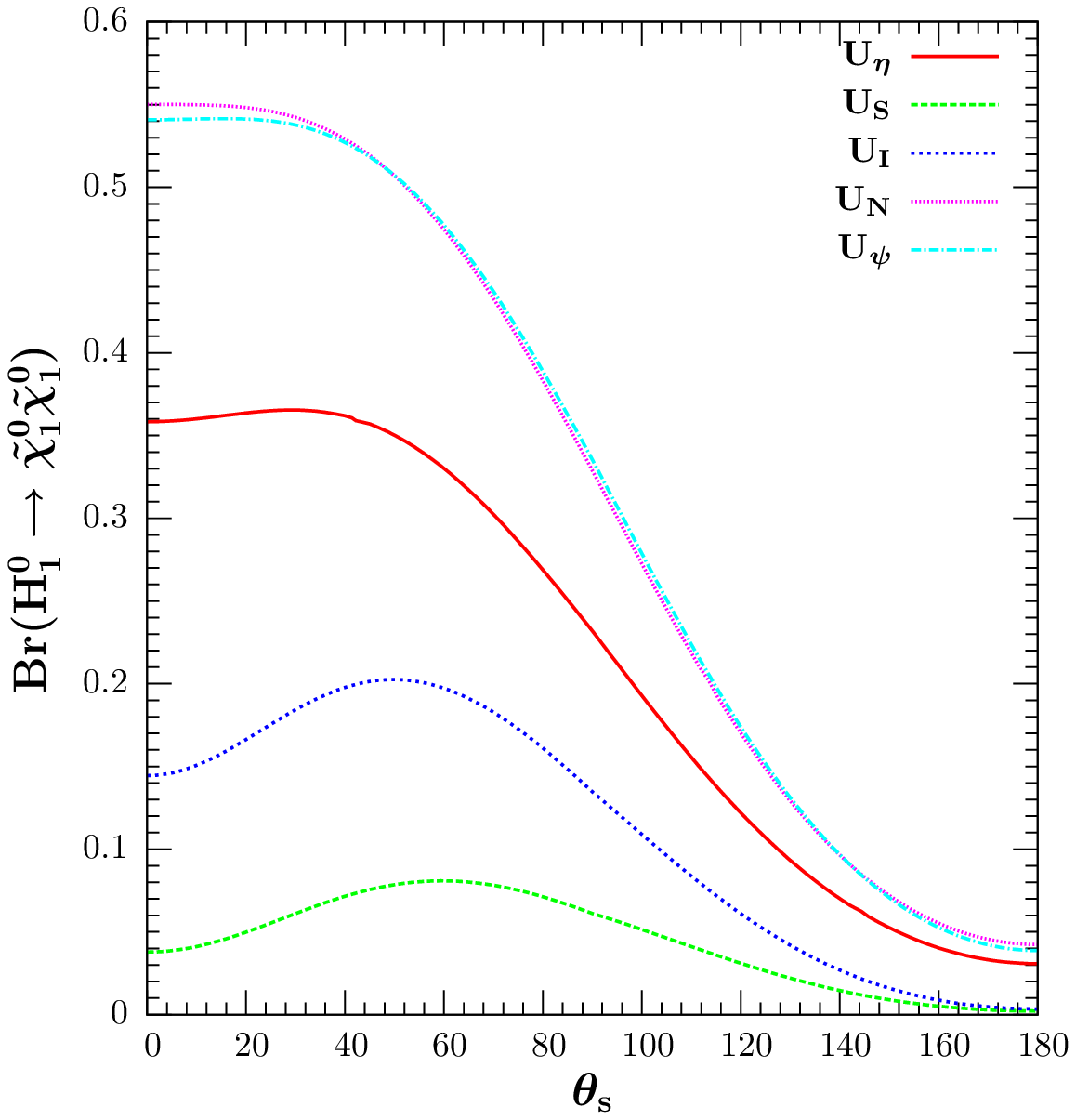}
      &\hspace*{-0.4cm}
         \includegraphics[width=2.5in,height=2.5in]{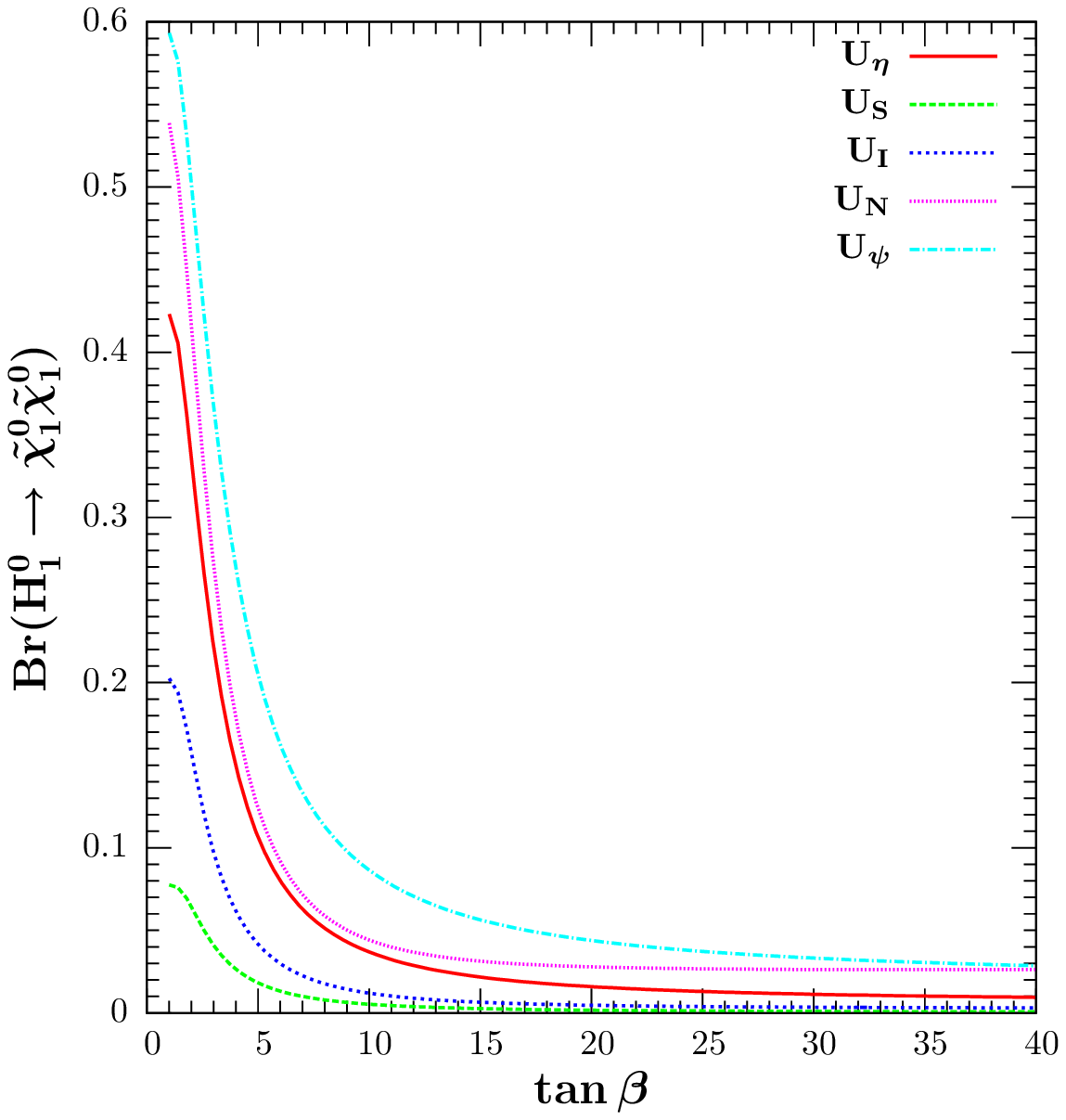}
&\hspace*{-0.4cm}
      \includegraphics[width=2.5in,height=2.5in]{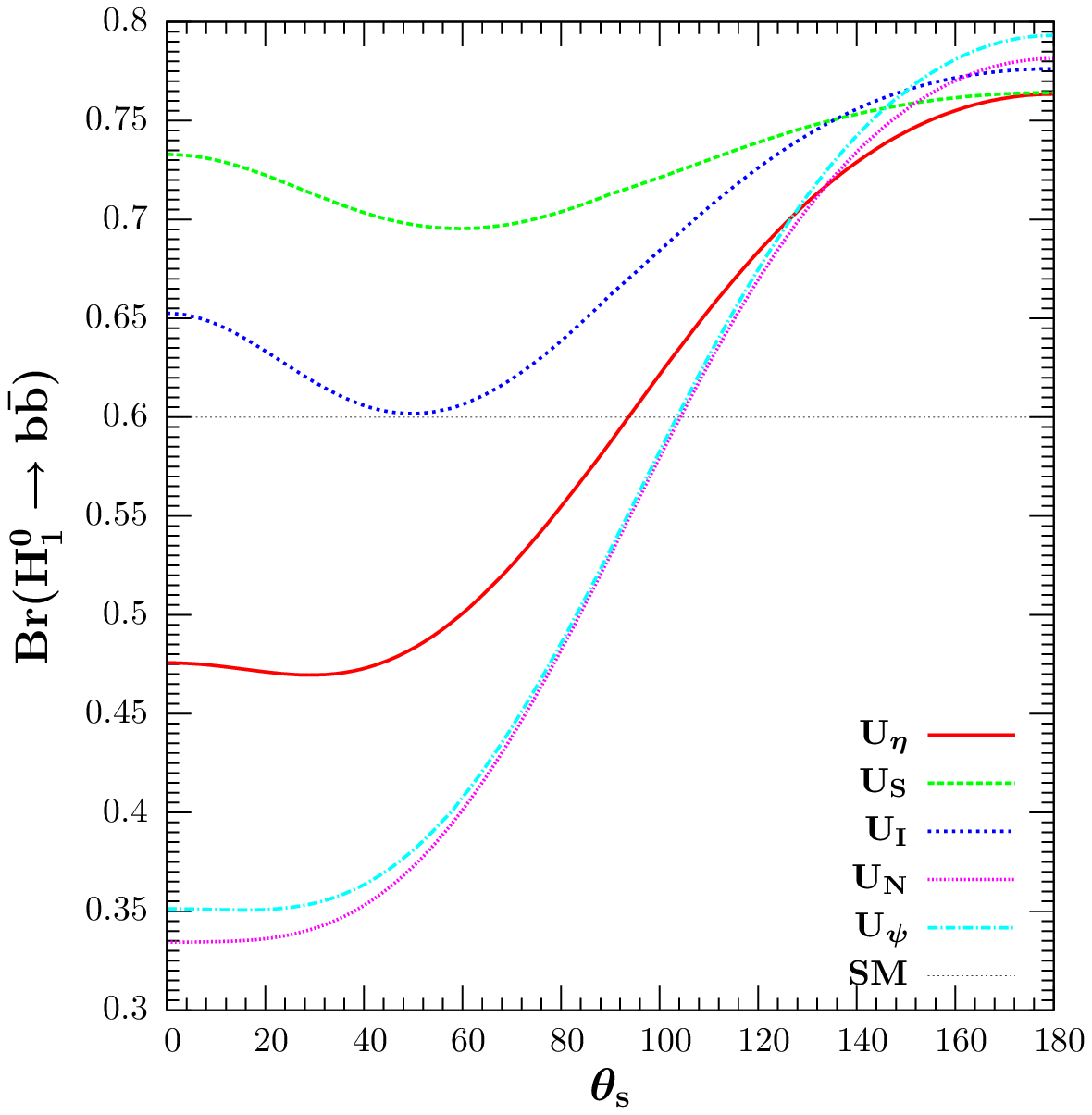}\\
\hspace*{-1.6cm}
               \includegraphics[width=2.5in,height=2.5in]{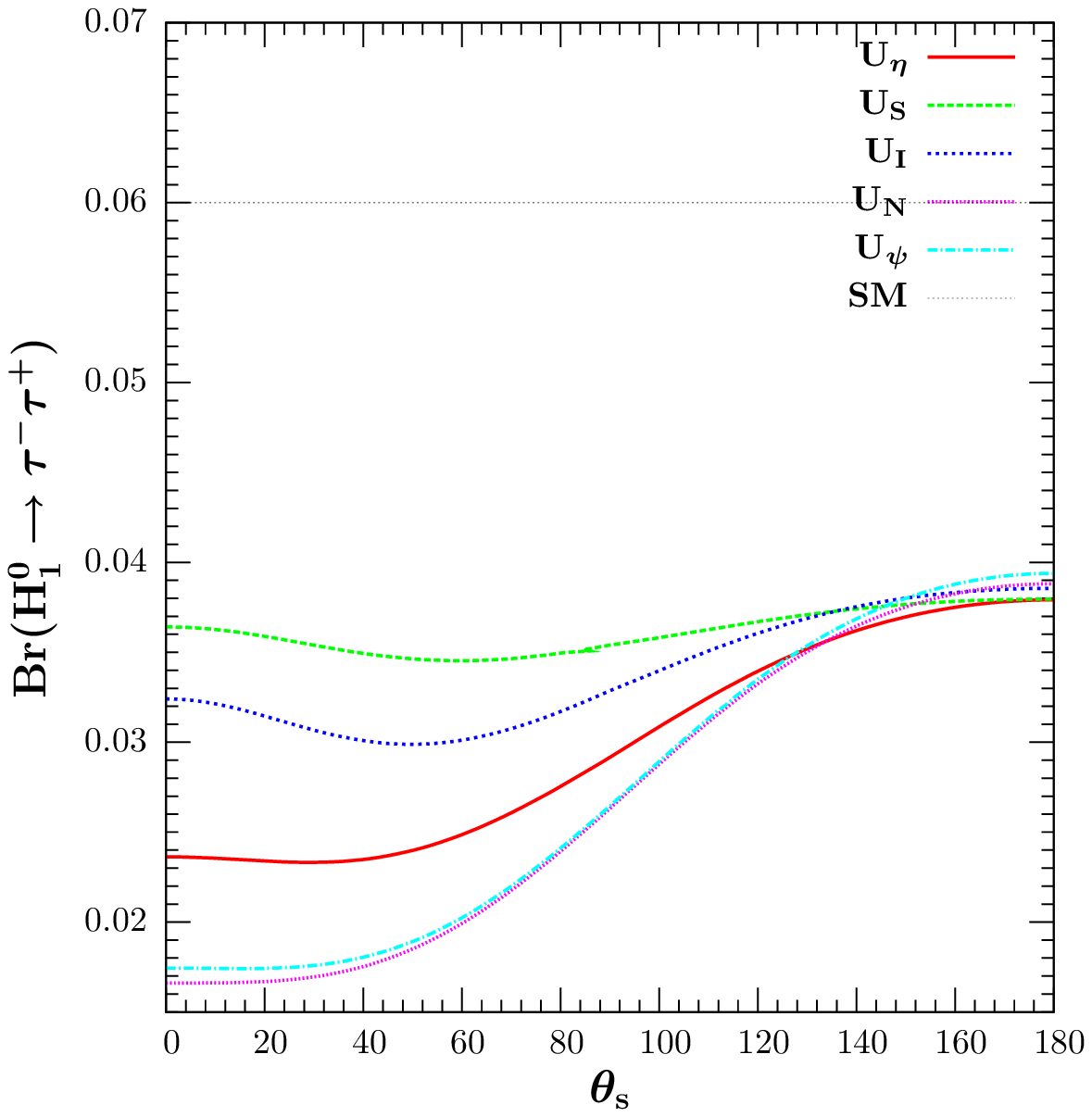}
   &\hspace*{-0.4cm}
              \includegraphics[width=2.6in,height=2.5in]{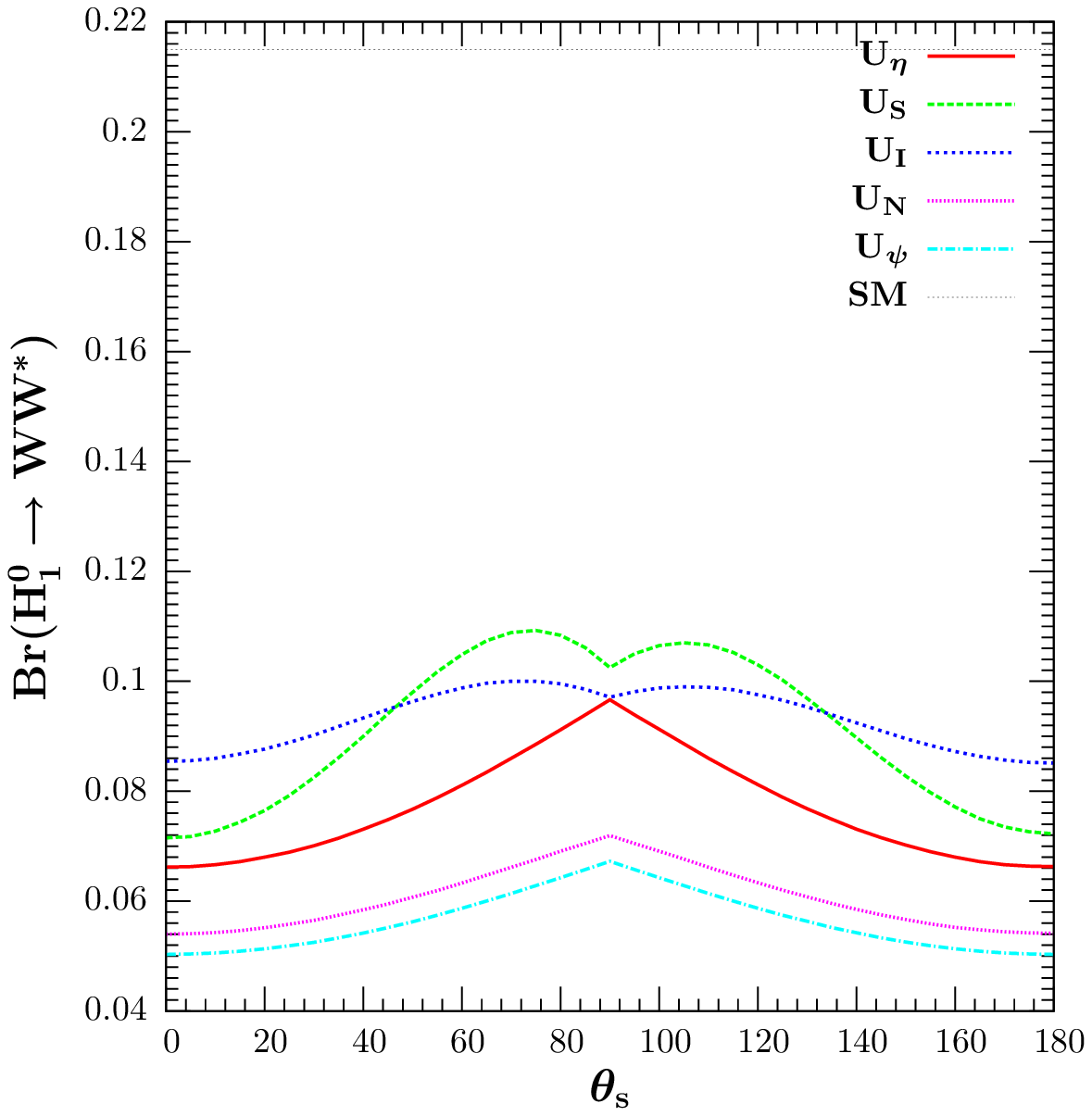}
  &\hspace*{-0.8cm}
               \includegraphics[width=2.7in,height=2.5in]{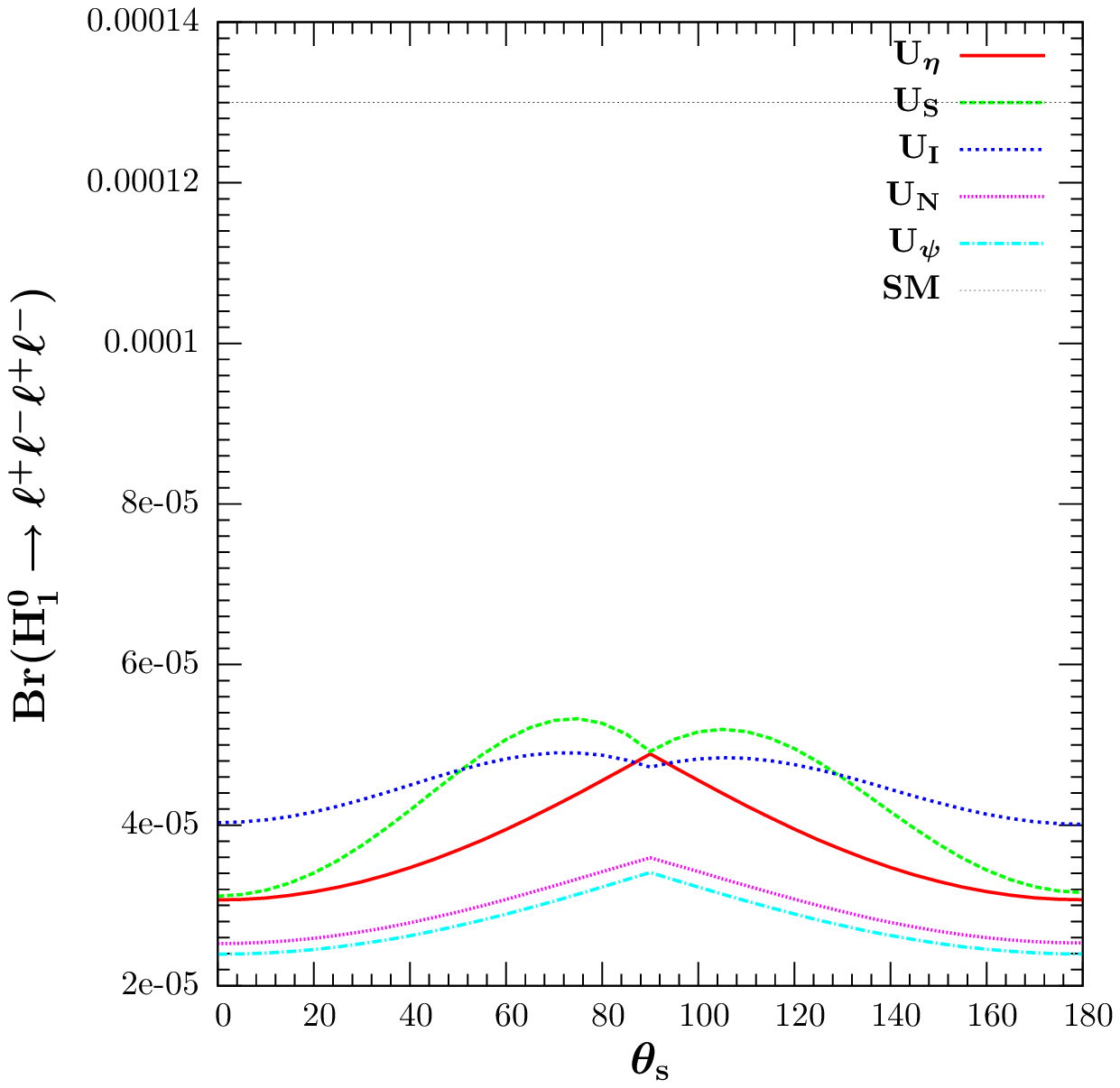}
\end{array}$
\end{center}
\vskip -0.3in
     \caption{BR$(H^0_1\rightarrow\chi^0_1\chi^0_1)$ as a  function of  {\sl\small $\theta_s$} and {\sl \small $\tan \beta$}, and BR$(H^0_1\rightarrow b\bar{b})$, BR$(H^0_1\rightarrow \tau^+ \tau^-)$, BR$(H^0_1\rightarrow WW^*)$ and BR$(H^0_1\rightarrow\ell^+\ell^-Z)$,  as functions of  {\sl\small $\theta_s$}   for the CP-violating versions of $U(1)_{\eta}$, $U(1)_S$, $U(1)_I$, $U(1)_N$ and
$U(1)_{\psi}$
    models. When available, we also show the value of the corresponding SM quantity.
} \label{fig:mH1decays}
\end{figure}

We list the
 dominant decay branching ratios (in \%) for the lightest neutral Higgs in our model and for comparison, in the SM in Table \ref{tab: cp brnch rat}, again for no CP violation ($\theta_s=0$) and  with CP violation (with phases as given in Table \ref{tab: cp inputs}). The branching ratios, as well as the cross sections are largely independent of the $\theta_t$ phase. One can see  that, while the production cross sections are fairly independent of the CP violating phase $\theta_s$,  the branching ratios are not, showing significant differences between the various $U(1)^\prime$ scenarios and the SM in the branching ratios. First, given the fact that the lightest neutralino (the LSP) has mass $m_{{\tilde \chi}^0_1} <m_{H^0_1}/2$, the Higgs boson has a considerable branching ratio into $\tilde\chi_{1}^0\tilde\chi_{1}^0$, that is, a significant invisible width\footnote{Note that in principle the Higgs boson can decay into sneutrinos, which can then cascade into neutralinos, contributing to the invisible width. We preclude this possibility here, as $m_{\tilde \nu}<m_{H_1^0}/2$ would require soft left- handed slepton masses of $\cal O$(100) GeV, in conflict with the EDM constraints.}. This is accompanied by a reduction in the branching ratio to other two-body decays, in particular $\tau^+ \tau^-$ and $WW^*$. Of all the $U(1)^\prime$ scenarios, the invisible width is the smallest in $U(1)_S$, though comparable with the decay width into $\tau^+ \tau^-$ for the case of no CP violation. For the other $U(1)^\prime$ models, the branching ratio for the invisible decay goes from a low 9\% in $U(1)_I$ with no CP violating phases, to 54\% in $U(1)_\psi$ with CP violation. A  general feature emerging from Table \ref{tab: cp brnch rat} is that the invisible width is enhanced in the presence of CP violation ($\theta_s \ne 0$) over the case with $\theta_s=0$. This is particularly strong in the case of $U(1)_S$, where the branching ratio increases for $\theta_s$  (as in Table \ref{tab: cp inputs})  to 3 times of its CP-conserving value, and for  $U(1)_I$  where it increases more than twofold. The decay into the invisible mode can reach over 50\%, which is similar to the value obtained in the MSSM \cite{AlbornozVasquez:2011aa}.  Note that the decay into invisible modes is sometimes at the expense of the main SM decay into $b {\bar b}$. In two of the models studied, $U(1)_S$ and $U(1)_I$ the $H_1^0 \to b {\bar b}$ branching ratio is in fact {\it increased} with respect to the SM value, while in $U(1)_\eta, U(1)_N$ and $U(1)_\psi$ it is suppressed with respect the SM expectations. But  a general feature of all  these models is the strong  suppression of the $H_1^0 \to (W^+ W^{*\,-}+W^{*\,+}W^-)$ and $H \to \tau^+ \tau^*$ decay modes, expected to have a branching ratio of $21.5$\% and 6\%, respectively, in the SM, but much smaller here. The branching ratio for the decay $H^0_1 \to \tau^+ \tau^-$ is between $ \sim 2-3.5\%$, while that  for $H_1^0 \to WW^*$  ranges between $\sim 5.5 -12\%$.  In a nutshell, the Higgs decay into the invisible mode ${\tilde \chi}_1^0 {\tilde \chi}_1^0$, is at the expense of $H_1^0 \to W^+ W^-$ and $\tau^+ \tau^-$ in {\it all} models, and occasionally due to a suppression of $H_1^0 \to b {\bar b}$ in {\it some} models. This behavior is not unexpected, as previous studies have indicated that for light Higgs masses, the decay into neutralinos and Higgs pseudoscalar pairs (if kinematically allowed) dominate, at the expense of the SM decay modes. Increasing the lightest Higgs mass opens allowed channels, but the branching ratios are affected by the mixing with the singlet Higgs field, the pseudoscalar  and the effect of the CP violating phase. However, due to  differences in decay patterns among various anomaly-free versions of the $U(1)^\prime$ models, a more precise measurement of the Higgs boson branching ratios at the LHC will serve not only to differentiate between the SM and the $U(1)^\prime $ model, but among the different  versions of $U(1)^\prime$'s.
 %%%%%%%%%%%%%%%%%%%%%%
%%%%%%%%%%%%%%%%%%%%%%
\setlength{\voffset}{-0.5in}
\begin{table}[htbp]
 \begin{center}
\setlength{\extrarowheight}{-5.8pt} \small
\begin{tabular*}{0.95\textwidth}{@{\extracolsep{\fill}} ccccccc}
\hline\hline
 $\rm Branching~ Ratio$  & $U(1)_{\eta}$  &$U(1)_{S}$ &$U(1)_{I}$&$U(1)_{N}$&$ U(1)_{\psi}$& SM
 \\ \cline{1-5}\cline{1-7}
$\rm BR(H^0_1\rightarrow\tilde\chi_{1}^0\tilde\chi_{1}^0)$&$36.0(34.0)$&$8.0(2.6)$&$20.0(9.0)$&$49.0(41.0)$&$54.0(42)$&$-$\\
$\rm BR(H^0_1\rightarrow b\bar{b})$&$48.0(49.0)$&$70.0(73.0)$&$60.0(66.0)$&$38.0(44.0)$&$36.0(43.0)$&60\\
$\rm BR(H^0_1\rightarrow \tau^- \tau^+)$&$2.3(2.4)$&$3.5(3.6)$&$3.0(3.3)$&$1.9(2.2)$&$1.8(2.2)$&6\\
$\rm BR(H^0_1\rightarrow  W W^*)$&$7.4(7.2)$&10.9$(11.1)$&9.8$(12.0)$&$6.1(7.5)$&$5.3(6.6)$&21.5\\
\hline\hline
\end{tabular*}
\caption{\sl\small Dominant branching ratios (in \%) of $H^0_1$ decay channels for the
CP-violating (CP-conserving) version of the $U(1)_{\eta}$, $U(1)_S$, $U(1)_I$, $U(1)_N$ and
$U(1)_{\psi}$
    scenarios considered,  and in the SM. } \label{tab: cp brnch rat}
\end{center}
 \end{table}
%%%%%%%%%%%%%%%%%%%%
 In Fig. \ref{fig:mH1decays} we plot the variation of the branching ratios of the lightest Higgs boson with the CP violating phase $\theta_s$.   In the first two panels, we depict the dependence of the BR$(H^0_1\rightarrow \chi^0_1\chi^0_1)$ with $\theta_s$ and $\tan \beta$. As we have seen previously $\tan \beta \sim 1-2$ (as in Table \ref{tab: cp inputs}), and in that region the invisible decay width is large, and very sensitive to $\tan \beta$. We show the variation of the branching ratios of the other dominant SM and $U(1)^\prime$ modes, as well as the that for BR($H_1^0 \to 4 \ell)$, because the LHC is sensitive to this decay in the 124-126 GeV mass range, and the value is expected to become more precise. Note that we did not include any of the loop dominated decays, such as $H_1^0 \to gg, \gamma \gamma$, as these are sensitive to the masses and mixing parameters of the (numerous) particles in the loop, and there no new contributions to these processes with respect to  MSSM. The third panel in the top row of the figure shows that, while the BR$(H_1^0 \to b {\bar b})$ in SM seems to fall somewhere in the middle of predictions for $U(1)^\prime$  models, the SM BR$(H_1^0 \to \tau^+ \tau^-)$ (bottom row, left side) is 6\%, and outside the range of $U(1)^\prime$ models, and  so is the SM value for the BR($H_1^0 \to WW^*)$ (bottom row, middle panel). BR$(H_1^0 \to 4 \ell)$ (bottom row, right panel) is also beyond the upper high end of $U(1)^\prime$ models predictions; the value expected in the SM is $0.013\%$ while in the $U(1)^\prime$ models,  the BR's fall in the $\sim 0.0025-0.0055\%$ range. The results for these decay widths might be more meaningful experimentally than the invisible Higgs width, which is difficult to measure.
 %%%%%%%%%%%%%%%%%%%%%%%%%%%%%%%%%%%%%%%%%%%%%%%%%%%%%%%%
%%%%%%%%%%%%%%%%%%%%%%%%%%%%%%%%%%%%%%%%%%%%%%%%%%%%%%%%%%%%%%%%%%%%%%%%%%%%%%%%%%%%%%%%%%%%%%%%%%

\subsection{ The second lightest  neutral Higgs boson}
\label{subsec:H2}

\begin{figure}[htb]
%\vskip -0.3in
\begin{center}$
 \begin{array}{cc}
         \hspace*{-0.6cm}
      \includegraphics[width=2.7in,height=2.5in]{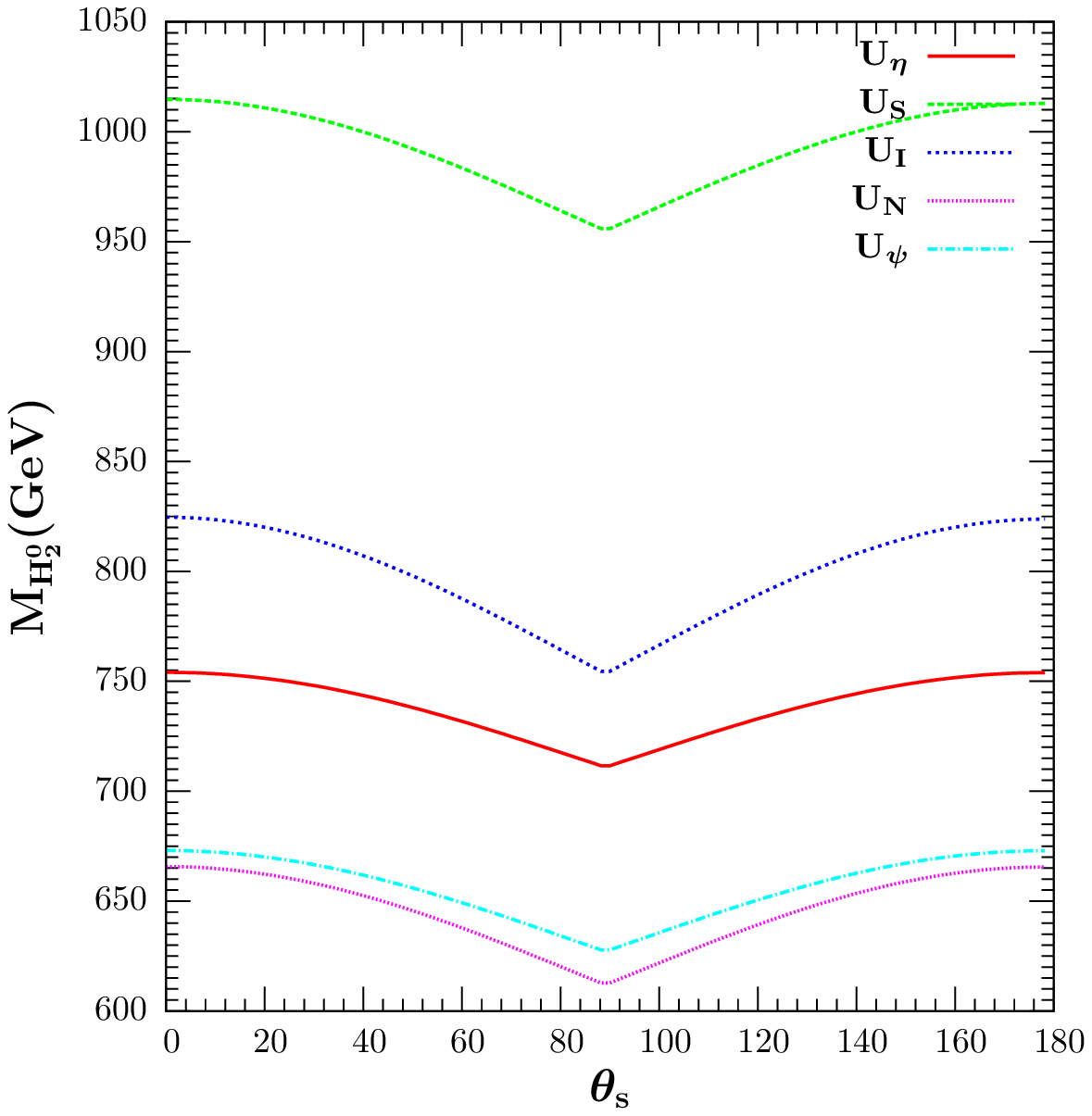}
\hspace*{1.0cm}
    \includegraphics[width=2.7in,height=2.5in]{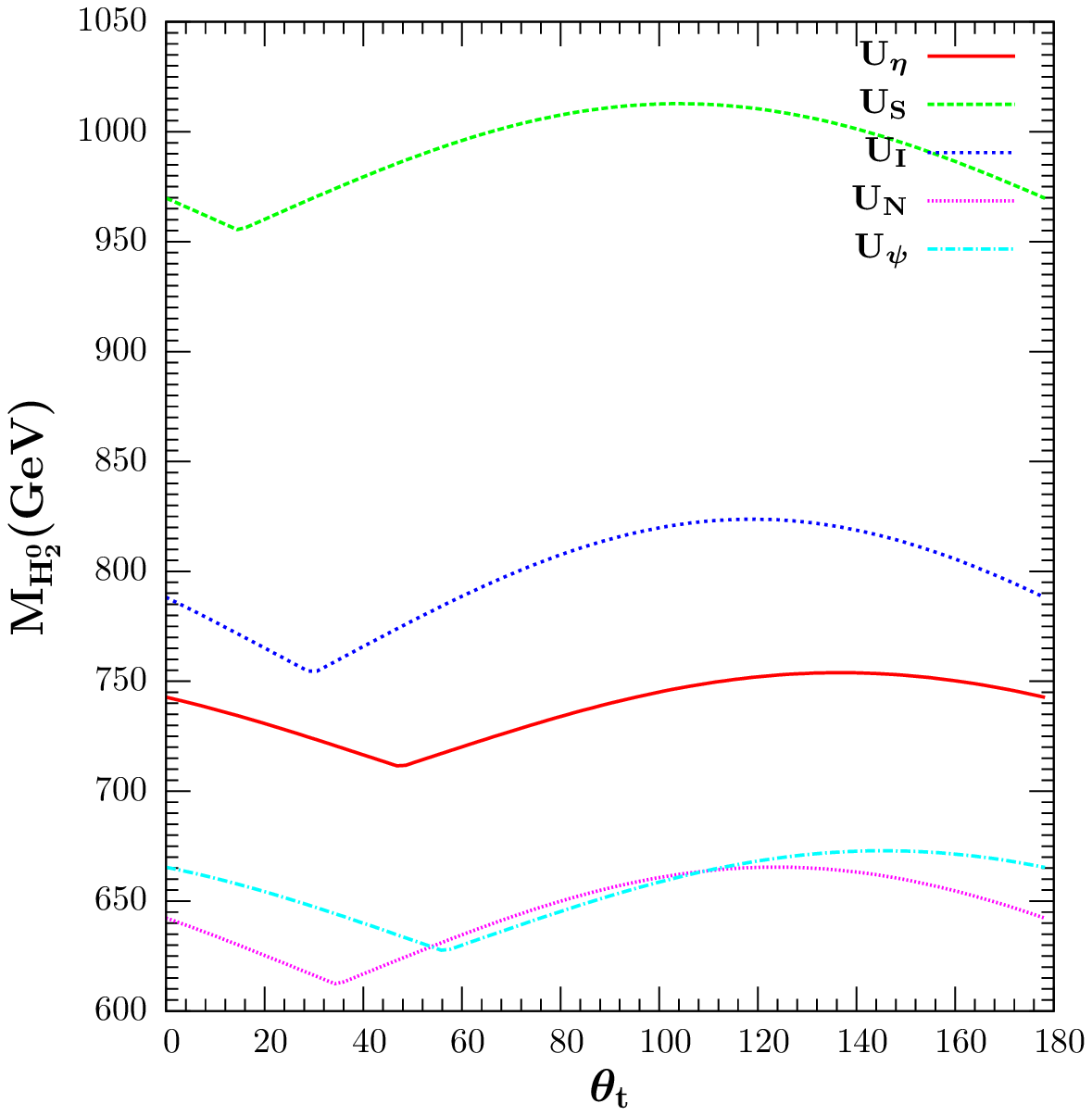}
\end{array}$
\end{center}
\vskip -0.3in
     \caption{ $M_{H^0_2}$ as a function of  \sl\small $\theta_s$ (the phase of
the new singlet S) and $\theta_t$ (the phase of the soft coupling
$A_t$)   for the CP-violating versions of the $U(1)_{\eta}$, $U(1)_S$, $U(1)_I$, $U(1)_N$ and
$U(1)_{\psi}$
   models.
} \label{fig:mH2}
\end{figure}

If the underlying symmetry in nature is not the SM, it is very likely that more Higgs boson states will be observed. The $U(1)^\prime$ models all predict additional neutral and charged Higgs states. The present collider bounds indicate that the mass of the second lightest Higgs boson must be heavier than about $600$ GeV. In our model, this mass shows explicit dependence on the CP violating phases $\theta_s$ and $\theta_t$. This dependence is correlated with the lightest boson mass. As the $m_{H}>600$ GeV mass region will be available to LHC working at increased $\sqrt{s}=14$ TeV, we show the mass dependence of the second lightest neutral Higgs boson in Fig. \ref{fig:mH2}.  The variation of this mass with either on the CP violating phases $\theta_s$ or $\theta_t$ is not as pronounced as for the lightest Higgs boson. Unlike in MSSM, in the majority of models under study this state appears to have a significant component of the pseudoscalar $A^0$. We leave the details of the decay width for later, when more experimental information could become available.
%%%%%%%%%%%%%%%%%%%%%%%%%%%%%%%%%%%%%%%%%%%%%%%%%%%%%%%%%%%
%%%%%%%%%%%%%%%%%%%%%%%%%%%%%%%%%%%%%%%%%%%%%%%%%%%%%%%%%%%%%
\newpage
\section{Discussion and Conclusion}
\label{sec:conclusion}

The recent discovery of a Higgs-like boson at the LHC does not preclude the possibility of beyond the Standard Model (BSM) physics. With increased energy and luminosity, the couplings of the Higgs boson to SM particles will be measured with increased precision.  In addition to the SM modes, the BSM Higgs boson can decay invisibly (to neutralinos, heavy neutrinos, or additional scalars). Our work investigates such a possibility, in a U(1)$^\prime$-extended supersymmetric model, by analyzing the decay patterns of the lightest neutral Higgs boson. This study is motivated by  the fact that the
composition of the Higgs bosons is different from one in the SM or MSSM
and hence, production and decay mechanisms are affected. Also significant is that  $U(1)^\prime$ models, unlike the SM, predict a light Higgs boson ($m_{H^0_1} \simeq 125$ GeV) naturally.

We chose anomaly-free versions of $U(1)^\prime$ motivated by breaking of string-inspired $E_6SSM$, and study the effects for both the CP-conserving and CP-violating scenarios, and compare the lightest Higgs boson production and decay to that in the SM. Our analysis has two goals: one is to analyze effects of CP violation on Higgs masses and decays, the other is look for differences among   each of the  $U(1)^\prime$ models for decay patterns, and identify characteristic signatures.

We perform a  complete study of Higgs sector of the
effective U(1)$^\prime$ models, starting with calculation of masses and mixings in the Higgs sector, and including corrections from the stop and sbottom sector to one-loop level. Then we introduce benchmark scenarios for each $E_6SSM$ motivated $U(1)^\prime$ model, defined in terms of soft parameters, and the Higgs, $Z^\prime$ and sparticle spectra obtained for the benchmarks. We include a complete spectrum for the neutralinos, and include the saturation of the relic density constraint for each of the five versions of the $U(1)^\prime$ models.  Our mass spectra calculation is restricted by the  inclusion of  all the known constraints on the low energy spectrum, and including all the recent constraints on the lightest Higgs boson mass, and also for rare decays and cosmological constraints.

We then investigate the cross sections in channels (the vector fusion channel and the associated Higgs production with a vector boson) most propitious to look for the Higgs boson to decay invisibly.  While the cross sections are not significantly affected by the CP phases (coming from the effective $\mu$ parameter and the scalar trilinear couplings), the masses and the branching ratios show significant variations.  With one exception, the decay into the lightest neutralino pair is significant in all, and dominant in two of the five $U(1)^\prime$ models under investigation. The invisible decay comes with, sometimes a  suppression of the $b{\bar b}$ decay mode, from 60\% to as low as 36\%,  except for $U(1)_S$ and $U(1)_I$, where the branching ratio is enhanced with respect to the SM, up to 73\%, in the absence of CP violating phases. All models exhibit a  strong suppression of $\tau^+ \tau^-$ mode (by a factor of 2-3), of $WW^*$ (by a factor of 2-5) and of $4\ell$  by the same factor. Some of these branching ratios seem to be in agreement with the present LHC data \cite{ATLASupdate,CMSupdate}, although the measurements are not yet precise enough for a conclusive statement.  The strong suppression in all $U(1)^\prime$ models  of the decay into $WW^*$ can be traced to the mixing with the singlet, the pseudoscalar, and the CP-phase contribution, all which are known to modify the couplings of the Higgs boson with respect to their SM values. Overall, we find that Higgs phenomenology in $U(1)^\prime$ model is significantly affected by the CP phases, especially $\theta_s$, and yields distinct signatures.  The resulting signatures are unlike those of the NMSSM with CP violation, where the branching ratios of the lightest neutral Higgs boson are fairly independent of the values of CP phase $\theta_s$ \cite{Ham:2007mt}. Some of the signals in $U(1)^\prime$ are typical of the anomaly-free versions of the models studied, others are characteristic  for a scenario (such as the enhancement of the branching ratio into $b {\bar b}$ in $U(1)_S$ and $U(1)_I$). While other generic tests of CP violation in the supersymmetric sector exist, such as measuring chargino polarization \cite{Frank:2006yh}, the dependence of the masses and decay patterns of the Higgs boson with the phases are a much more promising indications for  CP violation in $U(1)^\prime$.  Such signatures can be probed at the LHC, and are within reach at $\sqrt{s}=14$ TeV  with luminosity ${\cal L}=100~fb^{-1}$.

The decay patterns would enable to distinguish $U(1)^\prime $ models from the SM, but also from each other. For instance,  $U(1)_S$ and $U(1)_I$ show some similar decay patterns, insofar as the decay $H_1^0 \to b {\bar b}$ is  dominant.  Among all the models studied, $U(1)_S$ is the only one where the branching ratio of Higgs decay into neutralinos is  below 10\%; while in $U(1)_I$ the branching ratio into invisible modes is in the 10-20\% range. In $U(1)_\eta$, $U(1)_N$ and $U(1)_\psi$, the partial width into the invisible mode is significant, but in $U(1)_\eta$ it is still slightly below that into $b {\bar b}$. Distinguishing between $U(1)_N$ and $U(1)_\psi$ could also be based on the branching ratio into the invisible channel, which can be over  50\% in $U(1)_\psi$, but under 50\% in $U(1)_N$.

The characteristic signatures at the LHC would be distinctive kinematic distribution of the two quark jets in the Higgs production through vector boson fusion, compared to the $Zjj$ and $Wjj$ backgrounds. In the Higgs production with an associated vector boson, the $ZH$ associated production seems more promising, as a clean signal in the dilepton +$\EmissT$ channel will have little background, unlike the $WH$ model where the single lepton +$\EmissT$ suffers from large background effects from off-shell Drell-Yan production, as previously discussed in the literature \cite{Godbole:2003it}.
 This scenario also has consequences for other neutral Higgs states, and for the charged Higgs, the analyses of which await more data.

%%%%%%%%%%%%%%%%%%%%%%%%%%%%%%%%%%%%%%%%%%%%%%%%%%%

\section{Acknowledgments}

The work of  M.F.  is supported in part by NSERC under grant number
SAP105354. The research of L. S. is  supported in part by The
Council of Higher Education of Turkey (YOK).

%%%%%%%%%%%%%%%%%%%%%%%%%%%%%%%%%%%%%%%%%%%%%%%%%%
%\clearpage
%\section*{Appendix A: The Lagrangian}
%\setcounter{equation}{0}
\appendix

\section{Explicit Mass Formula}
\label{sec:appendix}

In these appendices we give the complete and detailed analytical expressions used in our calculations.

\subsection{Scalar Top and Scalar Bottom Masses}
Stop and sbottom mass-squared matrices show clearly the
differences between the MSSM and U(1)$^\prime$ extended models. As
can be seen from the following expressions, extra  charges and gauge
couplings affect LL and RR entries especially if the vacuum
expectation value of the $S$ field is sizable ($v_S\geq1$ TeV).

The entries of the field dependent ${M}^{2}$ for scalar top are
given by
\begin{eqnarray}
M_{LL}^{2}&=&M_{\widetilde{Q}}^{2}+Y_t^{2} |H_u|^2-\frac{1}{4} (g_2^2-\frac{g_Y^2}{3})(|H_u|^2 - |H_d|^2 )+g^2_{Y^\prime} {\cal Q}_Q({\cal Q}_u |H_u|^2 + {\cal Q}_d |H_d|^2+ {\cal Q}_S |S|^2 )\, , \nonumber\\
M_{RR}^{2}&=&M_{\widetilde{U}}^{2}+Y_t^{2} |H_u|^2-\frac{g_Y^2}{3}(|H_u|^2 - |H_d|^2 )+g^2_{Y^\prime} {\cal Q}_U({\cal Q}_u |H_u|^2+ {\cal Q}_d |H_d|^2 +{\cal Q}_S|S|^2 )\, , \nonumber\\
M_{LR}^{2}&=&M^{2\,\dagger}_{RL}=Y_t(A^*_t H_u^{0*}-Y_S S
H_d^{0}),
\end{eqnarray}
similarly for the scalar bottom mass-squared,  we have
\begin{eqnarray}
M_{LL}^{2}&=&M_{\widetilde{Q}}^{2}+Y_b^{2} |H_d|^2+\frac{1}{4} (g_2^2+\frac{g_Y^2}{3})(|H_u|^2 - |H_d|^2 )+g^2_{Y^\prime} {\cal Q}_Q({\cal Q}_u |H_u|^2 + {\cal Q}_d |H_d|^2+ {\cal Q}_S |S|^2 )\, , \nonumber\\
M_{RR}^{2}&=&M_{\widetilde{D}}^{2}+Y_b^{2} |H_u|^2+\frac{g_Y^2}{6}(|H_u|^2 - |H_d|^2 )+g^2_{Y^\prime} {\cal Q}_D({\cal Q}_u |H_u|^2 + {\cal Q}_d |H_d|^2+ {\cal Q}_S |S|^2 )\, , \nonumber\\
M_{LR}^{2}&=&M^{2\,\dagger}_{RL}=Y_b(A^*_b H_d^{0*}-Y_S S
H_u^{0}).
\end{eqnarray}

\subsection{Neutral Higgs Boson Masses}
The neutral Higgs masses are obtained by diagonalizing the $4 \times 4$ matrix in Eq. (\ref{higgsmassmat}). The explicit values of the entries are:
\begin{align}
\mathcal{M}^2_{1 1}&=\frac{\kappa}{3 \Sigma_t v_u} \bigg[\Sigma_t(3 \Delta_t^2(\Delta_b^2(2 Y_t^4 v_u^3 \ln(\frac{m_{\tilde{t}_1}^2 m_{\tilde{t}_2}^2}{m_t^4})+\mu  v_d(A_b C_b F_b Y_b^2+A_t C_t F_t Y_t^2))\nonumber\\
&+2 \mu^2 G_b Y_b^4 v_u(A_b C_b v_d-\mu  v_u)^2)+6 A_t^2 \Delta_b^2 G_t Y_t^4 v_u(\mu  C_t v_d-A_t v_u)^2+64 \pi^2 \Delta_b^2 \Delta_t^2 \lambda_u v_u^3)\nonumber\\
&+12 A_t \Delta_b^2(G_t-2) Y_t^4 \Delta_t^2 v_u^2(\mu  C_t v_d-A_t
v_u)+\mu  \chi  \Delta_b^2 v_d \Delta_t^2 \Sigma_t\bigg],
\end{align}
\begin{align}
\mathcal{M}^2_{1 2}&=\frac{ -\kappa}{3 \Sigma_b \Sigma_t} \bigg[\Delta_t^2 \Sigma_t(\Sigma_b(3 \mu  A_b Y_b^2(2 G_b Y_b^2(A_b C_b v_d-\mu  v_u)(A_b v_d-\mu  C_b v_u)+C_b \Delta_b^2 F_b)\nonumber\\
&-32 \pi^2 \Delta_b^2 v_d v_u \lambda_{\text{ud}})+6 \mu  \Delta_b^2(G_b-2) Y_b^4 v_d(\mu  v_u-A_b C_b v_d)+\mu  \chi  \Delta_b^2 \Sigma_b)\nonumber\\
&+6 \mu  \Delta_b^2 \Sigma_b Y_t^4(\mu  v_d-A_t C_t v_u)(A_t G_t \Sigma_t(\mu  C_t v_d-A_t v_u)+(G_t-2) \Delta_t^2 v_u)\nonumber\\
&+3 \mu  A_t \Delta_b^2 \Sigma_b C_t F_t Y_t^2 \Delta_t^2
\Sigma_t\bigg],
\end{align}
\begin{align}
\mathcal{M}^2_{1 3}&=\frac{\kappa }{3 v_S \Sigma_t} \bigg[\Delta_t^2 \Sigma_t(\Delta_b^2(3 \mu  F_b Y_b^2(2 \mu  v_u-A_b C_b v_d)+32 \pi^2 v_S^2 v_u \lambda_{\text{us}})\nonumber\\
&+6 \mu^2 G_b Y_b^4 v_u(A_b C_b v_d-\mu  v_u)^2-\mu  \chi  \Delta_b^2 v_d)-3 \mu  A_t \Delta_b^2 C_t v_d F_t Y_t^2 \Delta_t^2 \Sigma_t\nonumber\\
&-6 \mu  \Delta_b^2 v_d Y_t^4(\mu  v_d-A_t C_t v_u)(A_t G_t
\Sigma_t(\mu  C_t v_d-A_t v_u)+(G_t-2) \Delta_t^2 v_u)\bigg],
\end{align}
\begin{align}
\mathcal{M}^2_{1 4}&=\frac{2 \kappa \mu  \omega }{v_S \Sigma_t}  \bigg[\Sigma_t(\mu  A_b G_b Y_b^4 S_b \Delta_t^2(\mu  v_u-A_b C_b v_d)+A_t^2 \Delta_b^2 G_t Y_t^4 S_t(A_t v_u-\mu  C_t v_d))\nonumber\\
&-A_t \Delta_b^2(G_t-2) Y_t^4 S_t \Delta_t^2 v_u\bigg],
\end{align}
\begin{align}
\mathcal{M}^2_{2 2}&=\frac{ \kappa }{3 \Sigma_b v_d}\bigg[\Sigma_b(3 \Delta_t^2(\Delta_b^2(2 Y_b^4 v_d^3 \ln(\frac{m_{\tilde{b}_1}^2 m_{\tilde{b}_2}^2}{m_b^4})+\mu  v_u(A_b C_b F_b Y_b^2+A_t C_t F_t Y_t^2))\nonumber\\
&+2 A_b^2 G_b Y_b^4 v_d(A_b v_d-\mu  C_b v_u)^2)+6 \mu^2 \Delta_b^2 v_d G_t Y_t^4(\mu  v_d-A_t C_t v_u)^2+64 \pi^2 \Delta_b^2 \lambda_d v_d^3 \Delta_t^2)\nonumber\\
&-12 A_b \Delta_b^2(G_b-2) Y_b^4 v_d^2 \Delta_t^2(A_b v_d-\mu  C_b
v_u)+\mu  \chi  \Delta_b^2 \Sigma_b \Delta_t^2 v_u\bigg],
\end{align}
\begin{align}
\mathcal{M}^2_{2 3}&=\frac{\kappa}{3 \Sigma_b v_S}
\bigg[\Delta_t^2(\Sigma_b(3 \mu  \Delta_b^2(F_t Y_t^2(2 \mu v_d-A_t
C_t v_u)-A_b C_b F_b Y_b^2 v_u)\nonumber\\
&-6 \mu  A_b G_b Y_b^4 v_u(\mu v_u-A_b C_b v_d) (\mu  C_b v_u-A_b v_d)+32 \pi^2 \Delta_b^2 v_d \lambda_{\text{ds}} v_S^2)\nonumber\\
&+6 \mu  \Delta_b^2(G_b-2) Y_b^4 v_d v_u(A_b C_b v_d-\mu  v_u))+6 \mu^2 \Delta_b^2 \Sigma_b v_d G_t Y_t^4(\mu  v_d-A_t C_t v_u)^2\nonumber\\
&-\mu  \chi  \Delta_b^2 \Sigma_b \Delta_t^2 v_u\bigg],
\end{align}
\begin{align}
\mathcal{M}^2_{2 4}&=\frac{ 2 \kappa \mu  \omega }{\Sigma_b v_S} \bigg[\Sigma_b(A_b^2 G_b Y_b^4 S_b \Delta_t^2(A_b v_d-\mu  C_b v_u)+\mu  A_t \Delta_b^2 G_t Y_t^4 S_t(\mu  v_d-A_t C_t v_u))\nonumber\\
&-A_b \Delta_b^2(G_b-2) Y_b^4 S_b v_d \Delta_t^2\bigg],
\end{align}
\begin{align}
\mathcal{M}^2_{3 3}&=\frac{  \kappa }{3 v_s^2}\bigg[\Delta_t^2(3 \mu  v_u(\Delta_b^2 v_d(A_b C_b F_b Y_b^2+A_t C_t F_t Y_t^2)+2 \mu  G_b Y_b^4 v_u(A_b C_b v_d-\mu  v_u)^2\bigg]\\
&+64 \pi^2 \Delta_b^2 \lambda_s v_S^4)+6 \mu^2 \Delta_b^2 v_d^2 G_t
Y_t^4(\mu  v_d-A_t C_t v_u)^2+\mu  \chi  \Delta_b^2 v_d \Delta_t^2
v_u),
\end{align}
\begin{align}
\mathcal{M}^2_{3 4}&=\frac{2 \kappa \mu^2 \omega }{v_S^2}  \bigg[A_b
G_b Y_b^4 S_b \Delta_t^2 v_u(\mu  v_u-A_b C_b v_d)+A_t \Delta_b^2
v_d G_t Y_t^4 S_t(\mu  v_d-A_t C_t v_u)\bigg],
\end{align}
\begin{align}
\mathcal{M}^2_{4 4}&=\frac{ \kappa \mu  \omega^2}{3 v_d v_S^2 v_u} \bigg[3 \Delta_t^2(\Delta_b^2(A_b C_b F_b Y_b^2+A_t C_t F_t Y_t^2)+2 \mu  A_b^2 G_b Y_b^4 S_b^2 v_d v_u)\nonumber\\
&+6 \mu  A_t^2 \Delta_b^2 v_d G_t Y_t^4 S_t^2 v_u+\chi  \Delta_b^2
\Delta_t^2\bigg].
\end{align}

\subsection{CP-odd Tadpole Terms}
Explicit form of the CP-odd tadpole terms are
\begin{align}
&\mathcal{T}_{4}=\mu  A_S v_d \sin (\theta_{\Sigma }+\theta_S) +\frac{1}{32 \pi^2}  3 \mu  v_d (A_b F_b Y_b^2 S_b+A_t F_t Y_t^2 S_t), \nonumber\\
&\mathcal{T}_{5}=\mu  A_S v_u \sin (\theta_{\Sigma }+\theta_S) +\frac{1}{32 \pi^2}  3 \mu  v_u (A_b F_b Y_b^2 S_b+A_t F_t Y_t^2 S_t), \nonumber\\
&\mathcal{T}_{6}=\frac{\mu  A_S v_d v_u \sin (\theta_{\Sigma
}+\theta_S)}{v_S} +\frac{1}{32 \pi^2 v_S}  3 \mu  v_d v_u (A_b F_b
Y_b^2 S_b+A_t F_t Y_t^2 S_t).
\end{align}

\subsection{Charged Higgs Boson Masses}
Finally, the charged Higgs mass is obtained by diagonalizing the matrix in Eq. (\ref{higgsmassmat2}). One of the eigenvalues will be the Goldstone boson needed to give mass to the $W^\pm$ boson, the other is the real charged Higgs mass. The explicit entries in (\ref{higgsmassmat2}) are:
\begin{align}\mathcal{M}^{2\,\pm}_{1 1}&=\frac{1}{3 v^2 v_S^2 \Sigma_t
v_u} \bigg[ \kappa \Delta_b^2 v_d \Delta_t^2(\Sigma_t(\mu v_d^2
v_S^2 (3 A_b C_b F_b Y_b^2+\chi )+3 \mu A_b F_b Y_b^2 S_b v_S^2
v_u^2\nonumber\\
&+v_d v_u(8 \pi^2 v_d^2(g_2^2 v_S^2-4 \mu^2)-3 \mu^2 F_b Y_b^2
v_S^2))+3 Y_t^2 v_S^2(A_t F_t \Sigma_t (v_u(\mu S_t v_u-A_t
v_d)+\mu
C_t v_d^2)\nonumber\\
&-v_d \Sigma_t^2 v_u(F_t+G_t-2)+v_d(G_t-2) \Delta_t^2 v_u)+6 v_d
Y_t^4 v_S^2 \Sigma_t v_u^3(\ln (\frac{m_t^2}{Q^2})-1))\bigg],
\end{align}
\begin{align}\mathcal{M}^{2\, \pm}_{1 2}&=\frac{1}{3 v^2 \Sigma_b v_S^2} \bigg[ \kappa \Delta_b^2 \Delta
_t^2(\Sigma_b(3 \mu  A_t F_t Y_t^2 v_S^2(C_t v_u^2+v_d^2 S_t)+v_u(-3
\mu^2 v_d F_t Y_t^2 v_S^2\nonumber\\& +8 \pi^2 v_d v_u^2 (g_2^2
v_S^2-4 \mu ^2)+\mu  \chi v_S^2 v_u))+3 Y_b^2 v_S^2(A_b F_b
\Sigma_b(v_u(\mu C_b v_u-A_b v_d)+\mu S_b v_d^2)\nonumber\\&
-\Sigma_b^2 v_d v_u (F_b+G_b-2)+\Delta _b^2(G_b-2) v_d v_u)+6 Y_b^4
\Sigma_b v_d^3 v_S^2 v_u(\ln (\frac{m_b^2}{Q^2})-1))\bigg],
\end{align}
\begin{align}\mathcal{M}^{2\, \pm}_{2 1}&=\frac{1}{3 v^2 v_S^2 \Sigma_t} \bigg[ \kappa \Delta_b^2 \Delta
_t^2(\Sigma_t(\mu  v_d^2 v_S^2(3 A_b C_b F_b Y_b^2+\chi )+3 \mu A_b
F_b Y_b^2 S_b v_S^2 v_u^2\nonumber\\& +v_d v_u(8 \pi^2 v_d^2 (g_2^2
v_S^2-4 \mu ^2)-3 \mu^2 F_b Y_b^2 v_S^2))+3 Y_t^2 v_S^2(A_t F_t
\Sigma _t(v_u(\mu S_t v_u-A_t v_d)+\mu  C_t v_d^2)\nonumber\\&
-v_d\Sigma_t^2 v_u (F_t+G_t-2)+v_d(G_t-2) \Delta_t^2 v_u)+6 v_d
Y_t^4 v_S^2 \Sigma _t v_u^3(\ln(\frac{m_t^2}{Q^2})-1))\bigg],
\end{align}
\begin{align}\mathcal{M}^{2\, \pm}_{2 2}&=\frac{1}{3 v^2 \Sigma_b v_d v_S^2} \bigg[ \kappa \Delta_b^2
\Delta_t^2 v_u(\Sigma_b(3 \mu  A_t F_t Y_t^2 v_S^2(C_t v_u^2+v_d^2
S_t)+v_u(-3 \mu^2 v_d F_t Y_t^2 v_S^2\nonumber\\& +8 \pi^2 v_d v_u^2
(g_2^2 v_S^2-4 \mu^2)+\mu \chi v_S^2 v_u))+3 Y_b^2 v_S^2(A_b F_b
\Sigma _b(v_u(\mu  C_b v_u-A_b v_d)+\mu S_b
v_d^2)\nonumber\\
&-\Sigma_b^2 v_d v_u(F_b+G_b-2)+\Delta_b^2 (G_b-2) v_d v_u)+6 Y_b^4
\Sigma_b v_d^3 v_S^2 v_u (\ln(\frac{m_b^2}{Q^2})-1))\bigg].
\end{align}

\subsection{Auxiliary Expressions}
In the above expressions, we use the following short-hand notations:
\begin{align}
\chi=\sqrt{1024 \pi^4 A^2_S-9(A_b F_b Y^2_b S_b+A_t F_t Y^2_t
S_t)^2},
\end{align}
and
\begin{align}
&\lambda_u=\frac{1}{2} {\cal Q}_u^2 g^2_{Y^\prime}+\frac{g^2}{8}, \nonumber\\
&\lambda_d=\frac{1}{2} {\cal Q}_d^2 g^2_{Y^\prime}+\frac{g^2}{8}, \nonumber\\
&\lambda_s=\frac{1}{2} g^2_{Y^\prime} {\cal Q}_SS^2, \nonumber\\
&\lambda_{\text{ud}}={\cal Q}_d {\cal Q}_u g^2_{Y^\prime}-\frac{g^2}{4}+Y_S^2, \nonumber\\
&\lambda_{\text{ds}}={\cal Q}_d {\cal Q}_S g^2_{Y^\prime}+Y_S^2, \nonumber\\
&\lambda_{\text{us}}={\cal Q}_S {\cal Q}_u g^2_{Y^\prime}+Y_S^2. \nonumber\\
\end{align}


\begin{thebibliography}{99}

%\cite{:2012gk}
\bibitem{:2012gk}
  G.~Aad {\it et al.}  [ATLAS Collaboration],
  %``Observation of a new particle in the search for the Standard Model Higgs boson with the ATLAS detector at the LHC,''
  Phys.\ Lett.\ B {\bf 716}, 1 (2012).
 % [arXiv:1207.7214 [hep-ex]].
  %%CITATION = ARXIV:1207.7214;%%

%\cite{:2012gu}
\bibitem{:2012gu}
  S.~Chatrchyan {\it et al.}  [CMS Collaboration],
  %``Observation of a new boson at a mass of 125 GeV with the CMS experiment at the LHC,''
  Phys.\ Lett.\ B {\bf 716}, 30 (2012).
 % [arXiv:1207.7235 [hep-ex]].
  %%CITATION = ARXIV:1207.7235;%%

  %\cite{Ross:2012nr}
\bibitem{Ross:2012nr}
  G.~G.~Ross, K.~Schmidt-Hoberg and F.~Staub,
  %``The Generalised NMSSM at One Loop: Fine Tuning and Phenomenology,''
  JHEP {\bf 1208}, 074 (2012)
  [arXiv:1205.1509 [hep-ph]].
  %%CITATION = ARXIV:1205.1509;%%

  %\cite{Demir:2005ti}
\bibitem{Demir:2005ti}
  D.~A.~Demir, G.~L.~Kane and T.~T.~Wang,
  %``The minimal U(1)' extension of the MSSM,''
  Phys.\ Rev.\  D {\bf 72} (2005) 015012
  [arXiv:hep-ph/0503290].
  %%CITATION = PHRVA,D72,015012;%%


%\cite{Cvetic:1997ky}
\bibitem{Cvetic:1997ky}
M.~Cvetic, D.~A.~Demir, J.~R.~Espinosa, L.~L.~Everett and
P.~Langacker,
%``Electroweak breaking and the mu problem in supergravity models with an
%additional U(1),''
Phys.\ Rev.\ D {\bf 56}, 2861 (1997) [Erratum-ibid.\ D {\bf 58},
119905 (1998)] [hep-ph/9703317].




%\cite{muprob}
\bibitem{muprob}
%\bibitem{Kim:1983dt}
J.~E.~Kim and H.~P.~Nilles,
%``The Mu Problem And The Strong CP Problem,''
Phys.\ Lett.\ B {\bf 138}, 150 (1984);
%%CITATION = PHLTA,B138,150;%%
D.~Suematsu and Y.~Yamagishi,
%``Radiative symmetry breaking in a supersymmetric model with an extra U(1),''
Int.\ J.\ Mod.\ Phys.\ A {\bf 10}, 4521 (1995)
[arXiv:hep-ph/9411239];
%%CITATION = HEP-PH 9411239;%%
M.~Cvetic and P.~Langacker,
%``New Gauge Bosons from String Models,''
Mod.\ Phys.\ Lett.\ A {\bf 11}, 1247 (1996)
[arXiv:hep-ph/9602424];
%%CITATION = HEP-PH 9602424;%%
V.~Jain and R.~Shrock,
%``U(1)-A models of fermion masses without a mu problem,''
arXiv:hep-ph/9507238;
%%CITATION = HEP-PH 9507238;%%
Y.~Nir,
%``Gauge unification, Yukawa hierarchy and the mu problem,''
Phys.\ Lett.\ B {\bf 354}, 107 (1995) [arXiv:hep-ph/9504312].
%%CITATION = HEP-PH 9504312;%%

%\cite{Cvetic:1995rj}
\bibitem{Cvetic:1995rj}
  M.~Cvetic and P.~Langacker,
  %``Implications of Abelian Extended Gauge Structures From String Models,''
  Phys.\ Rev.\  D {\bf 54}, 3570 (1996)
  [arXiv:hep-ph/9511378].
  %%CITATION = PHRVA,D54,3570;%%
  %%CITATION = HEP-PH/9511378;%%


  %\cite{Babu:1996vt}
\bibitem{Babu:1996vt}
  K.~S.~Babu, C.~F.~Kolda and J.~March-Russell,
  %``Leptophobic U(1) $s$ and the R($b$) - R($c$) crisis,''
  Phys.\ Rev.\ D {\bf 54}, 4635 (1996)
  [hep-ph/9603212].
  %%CITATION = HEP-PH/9603212;%%

  %\cite{Cvetic:1996mf}
\bibitem{Cvetic:1996mf}
  M.~Cvetic and P.~Langacker,
  %``New Gauge Bosons from String Models,''
  Mod.\ Phys.\ Lett.\  A {\bf 11}, 1247 (1996)
  [arXiv:hep-ph/9602424].
  %%CITATION = MPLAE,A11,1247;%%


%\cite{Langacker:1999hs}
\bibitem{Langacker:1999hs}
  P.~Langacker, N.~Polonsky and J.~Wang,
  %``A Low-energy solution to the mu problem in gauge mediation,''
  Phys.\ Rev.\ D {\bf 60}, 115005 (1999)
  [hep-ph/9905252].
  %%CITATION = HEP-PH/9905252;%%


%\cite{Godbole:2003it}
\bibitem{Godbole:2003it}
  R.~M.~Godbole, M.~Guchait, K.~Mazumdar, S.~Moretti and D.~P.~Roy,
  %``Search for `invisible' Higgs signals at LHC via associated production with gauge bosons,''
  Phys.\ Lett.\ B {\bf 571}, 184 (2003)
  [hep-ph/0304137];
  %%CITATION = HEP-PH/0304137;%%
  %\cite{Ghosh:2012ep}
%\bibitem{Ghosh:2012ep}
  D.~Ghosh, R.~Godbole, M.~Guchait, K.~Mohan and D.~Sengupta,
  %``Looking for an Invisible Higgs Signal at the LHC,''
  arXiv:1211.7015 [hep-ph];
  %\cite{Eboli:2000ze}
%\bibitem{Eboli:2000ze}
  O.~J.~P.~Eboli and D.~Zeppenfeld,
  %``Observing an invisible Higgs boson,''
  Phys.\ Lett.\ B {\bf 495}, 147 (2000)
  [hep-ph/0009158].
  %%CITATION = HEP-PH/0009158;%%

  %\cite{Davoudiasl:2004aj}
\bibitem{Davoudiasl:2004aj}
  H.~Davoudiasl, T.~Han and H.~E.~Logan,
  %``Discovering an invisibly decaying Higgs at hadron colliders,''
  Phys.\ Rev.\ D {\bf 71}, 115007 (2005) [hep-ph/0412269];
  %%CITATION = HEP-PH/0412269;%%
%\cite{Frederiksen:1994me}
%\bibitem{Frederiksen:1994me}
S.~G.~Frederiksen, N.~Johnson, G.~L.~Kane and J.~Reid,
  %``Detecting invisible Higgs bosons at the CERN Large Hadron Collider,''
Phys.\ Rev.\ D {\bf 50}, 4244 (1994);
%%CITATION = PHRVA,D50,4244;%%
%\cite{Cao:2012im}
%\bibitem{Cao:2012im}
J.~-J.~Cao, Z.~Heng, J.~M.~Yang and J.~Zhu,
  %``Higgs decay to dark matter in low energy SUSY: is it detectable at the LHC ?,''
  JHEP {\bf 1206}, 145 (2012) [arXiv:1203.0694 [hep-ph]].
  %%CITATION = ARXIV:1203.0694;%%


  %\cite{Espinosa:2012vu}
\bibitem{Espinosa:2012vu}
  J.~R.~Espinosa, M.~Muhlleitner, C.~Grojean and M.~Trott,
  %``Probing for Invisible Higgs Decays with Global Fits,''
  JHEP {\bf 1209}, 126 (2012)
  [arXiv:1205.6790 [hep-ph]];
  %%CITATION = ARXIV:1205.6790;%%
  %\cite{Giardino:2012dp}
%\bibitem{Giardino:2012dp}
  P.~P.~Giardino, K.~Kannike, M.~Raidal and A.~Strumia,
  %``Is the resonance at 125 GeV the Higgs boson?,''
  Phys.\ Lett.\ B {\bf 718}, 469 (2012)
  [arXiv:1207.1347 [hep-ph]];
  %%CITATION = ARXIV:1207.1347;%%
  %\cite{Carmi:2012in}
%\bibitem{Carmi:2012in}
  D.~Carmi, A.~Falkowski, E.~Kuflik, T.~Volansky and J.~Zupan,
  %``Higgs After the Discovery: A Status Report,''
  JHEP {\bf 1210}, 196 (2012)
  [arXiv:1207.1718 [hep-ph]].
  %%CITATION = ARXIV:1207.1718;%%

  %\cite{Denner:2011mq}
\bibitem{Denner:2011mq}
  A.~Denner, S.~Heinemeyer, I.~Puljak, D.~Rebuzzi and M.~Spira,
  %``Standard Model Higgs-Boson Branching Ratios with Uncertainties,''
  Eur.\ Phys.\ J.\ C {\bf 71}, 1753 (2011)
  [arXiv:1107.5909 [hep-ph]].
  %%CITATION = ARXIV:1107.5909;%%
  %%CITATION = ARXIV:1211.7015;%%

    %\cite{Nakamura:2006ht}
\bibitem{Nakamura:2006ht}
  S.~Nakamura and D.~Suematsu,
  %``Supersymmetric extra U(1) models with a singlino dominated LSP,''
  Phys.\ Rev.\  D {\bf 75}, 055004 (2007)
  [arXiv:hep-ph/0609061].
  %%CITATION = PHRVA,D75,055004;%%

%\cite{Suematsu:2005bc}
\bibitem{Suematsu:2005bc}
  D.~Suematsu,
  %``Singlino dominating CDM in supersymmetric extra U(1) models,''
  Phys.\ Rev.\  D {\bf 73}, 035010 (2006)
  [arXiv:hep-ph/0511299].
  %%CITATION = PHRVA,D73,035010;%%


   % \cite{particledata}
  \bibitem{particledata}
  J. Beringer {\it et al}. (Particle Data Group), Phys. Rev. D {\bf 86}, 010001 (2012).


%\cite{Hall:2011au}
\bibitem{Hall:2011au}
  J.~P.~Hall, S.~F.~King, R.~Nevzorov, S.~Pakvasa and M.~Sher,
  %``Nonstandard Higgs Decays and Dark Matter in the E6SSM,''
  arXiv:1109.4972 [hep-ph].
  %%CITATION = ARXIV:1109.4972;%%



%\cite{Hall:2011zq}
\bibitem{Hall:2011zq}
  J.~P.~Hall and S.~F.~King,
  %``Bino Dark Matter and Big Bang Nucleosynthesis in the Constrained $E_6SSM with Massless Inert Singlinos,''
  JHEP {\bf 1106}, 006 (2011)
  [arXiv:1104.2259 [hep-ph]];
  %%CITATION = ARXIV:1104.2259;%%
  %\cite{Athron:2009cb}
%\bibitem{Athron:2009cb}
  P.~Athron, S.~F.~King, D.~J.~Miller, S.~Moretti and R.~Nevzorov,
  %``The constrained E(6)SSM,''
  PoS EPS {\bf -HEP2009}, 249 (2009)
  [arXiv:0910.0705 [hep-ph]];
  %%CITATION = ARXIV:0910.0705;%%
  %\cite{Athron:2011ew}
%\bibitem{Athron:2011ew}
  P.~Athron, J.~P.~Hall, S.~F.~King, S.~Moretti, D.~J.~Miller, R.~Nevzorov, S.~Pakvasa and M.~Sher,
  %``Collider phenomenology of the $E_6SSM$,''
  arXiv:1109.6373 [hep-ph].
  %%CITATION = ARXIV:1109.6373;%%

  %\cite{Frank:2006yh}
\bibitem{Frank:2006yh}
  M.~Frank, T.~Hahn, S.~Heinemeyer, W.~Hollik, H.~Rzehak and G.~Weiglein,
  %``The Higgs Boson Masses and Mixings of the Complex MSSM in the Feynman-Diagrammatic Approach,''
  JHEP {\bf 0702}, 047 (2007)
  [hep-ph/0611326];
  %%CITATION = HEP-PH/0611326;%%
  %\cite{Frank:2002qa}
%\bibitem{Frank:2002qa}
  M.~Frank, S.~Heinemeyer, W.~Hollik and G.~Weiglein,
  %``The Higgs boson masses of the complex MSSM: A Complete one loop calculation,''
  hep-ph/0212037;
  %%CITATION = HEP-PH/0212037;%%
  %\cite{Pilaftsis:1998dd}
%\bibitem{Pilaftsis:1998dd}
%\cite{Heinemeyer:2007aq}
%\bibitem{Heinemeyer:2007aq}
  S.~Heinemeyer, W.~Hollik, H.~Rzehak and G.~Weiglein,
  %``The Higgs sector of the complex MSSM at two-loop order: QCD contributions,''
  Phys.\ Lett.\ B {\bf 652}, 300 (2007)
  [arXiv:0705.0746 [hep-ph]];
  %%CITATION = ARXIV:0705.0746;%%
  A.~Pilaftsis,
  %``Higgs scalar - pseudoscalar mixing in the minimal supersymmetric standard model,''
  Phys.\ Lett.\ B {\bf 435}, 88 (1998)
  [hep-ph/9805373];
  %%CITATION = HEP-PH/9805373;%%
  %\cite{Pilaftsis:1999qt}
%\bibitem{Pilaftsis:1999qt}
  A.~Pilaftsis and C.~E.~M.~Wagner,
  %``Higgs bosons in the minimal supersymmetric standard model with explicit CP violation,''
  Nucl.\ Phys.\ B {\bf 553}, 3 (1999)
  [hep-ph/9902371];
  %%CITATION = HEP-PH/9902371;%%
  %\cite{Demir:1999hj}
%\bibitem{Demir:1999hj}
  D.~A.~Demir,
  %``Effects of the supersymmetric phases on the neutral Higgs sector,''
  Phys.\ Rev.\ D {\bf 60}, 055006 (1999)
  [hep-ph/9901389];
  %%CITATION = HEP-PH/9901389;%%
 %\cite{Choi:2000wz}
%\bibitem{Choi:2000wz}
  S.~Y.~Choi, M.~Drees and J.~S.~Lee,
  %``Loop corrections to the neutral Higgs boson sector of the MSSM with explicit CP violation,''
  Phys.\ Lett.\ B {\bf 481}, 57 (2000)
  [hep-ph/0002287];
  %%CITATION = HEP-PH/0002287;%%
  %\cite{Carena:2000yi}
%\bibitem{Carena:2000yi}
  M.~S.~Carena, J.~R.~Ellis, A.~Pilaftsis and C.~E.~M.~Wagner,
  %``Renormalization group improved effective potential for the MSSM Higgs sector with explicit CP violation,''
  Nucl.\ Phys.\ B {\bf 586}, 92 (2000)
  [hep-ph/0003180];
  %%CITATION = HEP-PH/0003180;%%
  {\it ibid.},
  %``Higgs boson pole masses in the MSSM with explicit CP violation,''
  Nucl.\ Phys.\ B {\bf 625}, 345 (2002)
  [hep-ph/0111245];
  %%CITATION = HEP-PH/0111245;%%
  %\cite{Ibrahim:2000qj}
%\bibitem{Ibrahim:2000qj}
  T.~Ibrahim and P.~Nath,
  %``Corrections to the Higgs boson masses and mixings from chargino, W and charged Higgs exchange loops and large CP phases,''
  Phys.\ Rev.\ D {\bf 63}, 035009 (2001)
  [hep-ph/0008237];
  %%CITATION = HEP-PH/0008237;%%
  {\it ibid.},
  %``Neutralino exchange corrections to the Higgs boson mixings with explicit CP violation,''
  Phys.\ Rev.\ D {\bf 66}, 015005 (2002)
  [hep-ph/0204092];
  %%CITATION = HEP-PH/0204092;%
  %\cite{Okada:1990vk}
  %\bibitem{Okada:1990vk}
  Y.~Okada, M.~Yamaguchi and T.~Yanagida,
  %``Upper bound of the lightest Higgs boson mass in the minimal supersymmetric standard model,''
  Prog.\ Theor.\ Phys.\  {\bf 85}, 1 (1991);
  %\cite{Ellis:1990nz}
%\bibitem{Ellis:1990nz}
  J.~R.~Ellis, G.~Ridolfi and F.~Zwirner,
  %``Radiative corrections to the masses of supersymmetric Higgs bosons,''
  Phys.\ Lett.\ B {\bf 257}, 83 (1991);
  %%CITATION = PHLTA,B257,83;%
%\bibitem{Haber:1990aw}
  H.~E.~Haber and R.~Hempfling,
  %``Can the mass of the lightest Higgs boson of the minimal supersymmetric model be larger than m(Z)?,''
  Phys.\ Rev.\ Lett.\  {\bf 66}, 1815 (1991);
  %%CITATION = PRLTA,66,1815;%%\cite{Ham:2002ps}
%\bibitem{Ham:2002ps}
  S.~W.~Ham, S.~K.~Oh, E.~J.~Yoo, C.~M.~Kim and D.~Son,
  %``Neutral Higgs boson masses of the MSSM at the one loop level in an explicit CP violation scenario,''
  Phys.\ Rev.\ D {\bf 68}, 055003 (2003)
  [hep-ph/0205244].
  %%CITATION = HEP-PH/0205244;%%



%\cite{Graf:2012hh}
\bibitem{Graf:2012hh}
  T.~Graf, R.~Grober, M.~Muhlleitner, H.~Rzehak and K.~Walz,
  %``Higgs Boson Masses in the Complex NMSSM at One-Loop Level,''
  JHEP {\bf 1210}, 122 (2012)
  [arXiv:1206.6806 [hep-ph]];
  %%CITATION = ARXIV:1206.6806;%%
  %\bibitem{Cheung:2010ba}
  K.~Cheung, T.~-J.~Hou, J.~S.~Lee and E.~Senaha,
  %``The Higgs Boson Sector of the Next-to-MSSM with CP Violation,''
  Phys.\ Rev.\ D {\bf 82}, 075007 (2010)
  [arXiv:1006.1458 [hep-ph]];
  %%CITATION = ARXIV:1006.1458;%%
  %\cite{Funakubo:2004ka}
%\bibitem{Funakubo:2004ka}
  K.~Funakubo and S.~Tao,
  %``The Higgs sector in the next-to-MSSM,''
  Prog.\ Theor.\ Phys.\  {\bf 113}, 821 (2005)
  [hep-ph/0409294];
  %%CITATION = HEP-PH/0409294;%%
  %\cite{Ham:2001wt}
%\bibitem{Ham:2001wt}
  S.~W.~Ham, S.~K.~Oh and D.~Son,
  %``Neutral Higgs sector of the next-to-minimal supersymmetric standard model with explicit CP violation,''
  Phys.\ Rev.\ D {\bf 65}, 075004 (2002)
  [hep-ph/0110052];
  %%CITATION = HEP-PH/0110052;%%
  %\cite{Ham:2003jf}
%\bibitem{Ham:2003jf}
  S.~W.~Ham, Y.~S.~Jeong and S.~K.~Oh,
  %``Radiative CP violation in the Higgs sector of the next-to-minimal supersymmetric model,''
  hep-ph/0308264.
  %%CITATION = HEP-PH/0308264;%%

    %\cite{Demir:2003ke}
\bibitem{Demir:2003ke}
  D.~A.~Demir and L.~L.~Everett,
  %``CP violation in supersymmetric U(1)' models,''
  Phys.\ Rev.\  D {\bf 69}, 015008 (2004)
  [arXiv:hep-ph/0306240].
  %%CITATION = PHRVA,D69,015008;%%
  %\cite{Cheung:2010ba}





    %\cite{Branco:2000dq}
\bibitem{Branco:2000dq}
  G.~C.~Branco, F.~Kruger, J.~C.~Romao and A.~M.~Teixeira,
  %``Spontaneous CP violation in the next-to-minimal supersymmetric standard model revisited,''
  JHEP {\bf 0107}, 027 (2001)
  [hep-ph/0012318];
  %%CITATION = HEP-PH/0012318;%%
  %\cite{Hugonie:2003yu}
%\bibitem{Hugonie:2003yu}
  C.~Hugonie, J.~C.~Romao and A.~M.~Teixeira,
  %``Spontaneous CP violation in nonminimal supersymmetric models,''
  JHEP {\bf 0306}, 020 (2003)
  [hep-ph/0304116].
  %%CITATION = HEP-PH/0304116;%%

  %\cite{Demir:2005kg}
\bibitem{Demir:2005kg}
  D.~A.~Demir, L.~Solmaz and S.~Solmaz,
  %``LEP indications for two light Higgs bosons and U(1)' model,''
  Phys.\ Rev.\  D {\bf 73}, 016001 (2006)
  [arXiv:hep-ph/0512134].

  %\cite{Coleman:1973jx}
\bibitem{Coleman:1973jx}
  S.~R.~Coleman and E.~J.~Weinberg,
  %``Radiative Corrections as the Origin of Spontaneous Symmetry Breaking,''
  Phys.\ Rev.\ D {\bf 7}, 1888 (1973).
  %%CITATION = PHRVA,D7,1888;%%

  % \cite{Demir:2010is}
   \bibitem{Demir:2010is}
  D.~A.~Demir, M.~Frank, L.~Selbuz and I.~Turan,
  %``Scalar Neutrinos at the LHC,''
  Phys.\ Rev.\  D {\bf 83}, 095001 (2011).
  %[arXiv:1012.5105 [hep-ph]].
  %%CITATION = PHRVA,D83,095001;%%



  %\cite{Choi:2006fz}
\bibitem{Choi:2006fz}
  S.~Y.~Choi, H.~E.~Haber, J.~Kalinowski and P.~M.~Zerwas,
  %``The neutralino sector in the U(1)-extended supersymmetric standard model,''
  Nucl.\ Phys.\  B {\bf 778} (2007) 85
  [arXiv:hep-ph/0612218].



 %\cite{Ibrahim:1997gj}
\bibitem{Ibrahim:1997gj}
  T.~Ibrahim and P.~Nath,
  %``The Neutron and the electron electric dipole moment in N=1 supergravity unification,''
  Phys.\ Rev.\ D {\bf 57}, 478 (1998)
  [Erratum-ibid.\ D {\bf 58}, 019901 (1998)]
  [Erratum-ibid.\ D {\bf 60}, 079903 (1999)]
  [Erratum-ibid.\ D {\bf 60}, 119901 (1999)]
  [hep-ph/9708456];
  %%CITATION = HEP-PH/9708456;%%
  %\cite{Abel:2001vy}
%\bibitem{Abel:2001vy}
  S.~Abel, S.~Khalil and O.~Lebedev,
  %``EDM constraints in supersymmetric theories,''
  Nucl.\ Phys.\  B {\bf 606} (2001) 151
  [arXiv:hep-ph/0103320].


%\cite{Regan:2002ta}
\bibitem{Regan:2002ta}
  B.~C.~Regan, E.~D.~Commins, C.~J.~Schmidt and D.~DeMille,
  %``New limit on the electron electric dipole moment,''
  Phys.\ Rev.\ Lett.\  {\bf 88} (2002) 071805.
  %%CITATION = PRLTA,88,071805;%%
%\cite{Baker:2006ts}
%\bibitem{Baker:2006ts}
  C.~A.~Baker {\it et al.},
  %``An improved experimental limit on the electric dipole moment of the
  %neutron,''
  Phys.\ Rev.\ Lett.\  {\bf 97}, 131801 (2006)
  [arXiv:hep-ex/0602020].
  %%CITATION = PRLTA,97,131801;%%


  %SPS1a'
\bibitem{AguilarSaavedra:2005pw}
  J.~A.~Aguilar-Saavedra {\it et al.},
  %``Supersymmetry parameter analysis: SPA convention and project,''
  Eur.\ Phys.\ J.\  C {\bf 46}, 43 (2006)
  [arXiv:hep-ph/0511344].
  %%CITATION = EPHJA,C46,43;%%

   %\cite{Langacker:2008yv}
\bibitem{Langacker:2008yv}
  P.~Langacker,
  %``The Physics of Heavy $Z^\prime$ Gauge Bosons,''
  Rev.\ Mod.\ Phys.\  {\bf 81}, 1199 (2009)
  [arXiv:0801.1345 [hep-ph]].
  %%CITATION = ARXIV:0801.1345;%%


  %U(1)_\psi
  %\cite{Langacker:1998tc}
\bibitem{Langacker:1998tc}
  P.~Langacker and J.~Wang,
  %``U(1)-prime symmetry breaking in supersymmetric E(6) models,''
  Phys.\ Rev.\ D {\bf 58}, 115010 (1998)
  [hep-ph/9804428];
  %%CITATION = HEP-PH/9804428;%%
 R.~N.~ Mohapatra, {\it Unification and supersymmetry: the
frontiers of quark-lepton physics} (2003) (Ed. Springer, Berlin).


  %U(1)_\chi
  %\cite{Hewett:1988xc}
\bibitem{Hewett:1988xc}
  J.~L.~Hewett and T.~G.~Rizzo,
  %``Low-Energy Phenomenology of Superstring Inspired E(6) Models,''
  Phys.\ Rept.\  {\bf 183}, 193 (1989).
  %%CITATION = PRPLC,183,193;%%

  %U(1)_eta
  %\cite{Witten:1985xc}
\bibitem{Witten:1985xc}
  E.~Witten,
  %``Symmetry Breaking Patterns in Superstring Models,''
  Nucl.\ Phys.\ B {\bf 258}, 75 (1985).
  %%CITATION = NUPHA,B258,75;%%


  %U(1)_S
%\cite{Erler:2002pr}
\bibitem{Erler:2002pr}
  J.~Erler, P.~Langacker and T.~-j.~Li,
  %``The $Z$ - $Z^\prime$ mass hierarchy in a supersymmetric model with a secluded U(1) -prime breaking sector,''
  Phys.\ Rev.\ D {\bf 66}, 015002 (2002)
  [hep-ph/0205001].
  %%CITATION = HEP-PH/0205001;%%

  %U(1)_I
  %\cite{Robinett:1982tq}
\bibitem{Robinett:1982tq}
  R.~W.~Robinett and J.~L.~Rosner,
  %``Mass Scales In Grand Unified Theories,''
  Phys.\ Rev.\  D {\bf 26}, 2396 (1982).

%U(1)_N
 %\cite{Barger:2003zh}
\bibitem{Barger:2003zh}
  V.~Barger, P.~Langacker and H.~-S.~Lee,
  %``Primordial nucleosynthesis constraints on $Z^\prime$ properties,''
  Phys.\ Rev.\ D {\bf 67}, 075009 (2003)
  [hep-ph/0302066];
  %%CITATION = HEP-PH/0302066;%%
  %\cite{Kang:2004bz}
%\bibitem{Kang:2004bz}
  J.~Kang and P.~Langacker,
  %``$Z$ ' discovery limits for supersymmetric E(6) models,''
  Phys.\ Rev.\ D {\bf 71}, 035014 (2005)
  [hep-ph/0412190];
  %%CITATION = HEP-PH/0412190;%%
   %%CITATION = HEP-PH/0510419;%%
  %\cite{Ma:1995xk}
%\bibitem{Ma:1995xk}
  E.~Ma,
  %``Neutrino masses in an extended gauge model with E(6) particle content,''
  Phys.\ Lett.\ B {\bf 380}, 286 (1996)
  [hep-ph/9507348].
  %%CITATION = HEP-PH/9507348;%%

  %\cite{King:2005jy}
\bibitem{King:2005jy}
  S.~F.~King, S.~Moretti and R.~Nevzorov,
  %``Theory and phenomenology of an exceptional supersymmetric standard
  %model,''
  Phys.\ Rev.\  D {\bf 73} (2006) 035009
  [arXiv:hep-ph/0510419].


%\cite{Belanger:2008sj}
\bibitem{Belanger:2008sj}
  G.~Belanger, F.~Boudjema, A.~Pukhov {\it et al.},
  %``Dark matter direct detection rate in a generic model with micrOMEGAs_2.2,''
  Comput.\ Phys.\ Commun.\  {\bf 180}, 747-767 (2009)
  [arXiv:0803.2360 [hep-ph]];
%\cite{Belanger:2010gh}
%\bibitem{Belanger:2010gh}
  G.~Belanger, F.~Boudjema, P.~Brun {\it et al.},
  %``Indirect search for dark matter with micrOMEGAs2.4,''
  [arXiv:1004.1092 [hep-ph]].

  \bibitem{calchep}
See the URL: http://theory.sinp.msu.ru/~pukhov/calchep.html;
%\cite{Pukhov:2004ca}
%\bibitem{Pukhov:2004ca}
  A.~Pukhov,
  %``CalcHEP 3.2: MSSM, structure functions, event generation, batchs, and
  %generation of matrix elements for other packages,''
  [arXiv:hep-ph/0412191].
  %%CITATION = HEP-PH/0412191;%%


\bibitem{wmap}
E.~Komatsu {\it et al.} [WMAP Collaboration],
  %``Seven-Year Wilkinson Microwave Anisotropy Probe (WMAP) Observations:
  %Cosmological Interpretation,''
  arXiv:1001.4538 [astro-ph.CO].
  %%CITATION = ARXIV:1001.4538;%%

%\cite{Spergel:2006hy}
\bibitem{Spergel:2006hy}
  D.~N.~Spergel {\it et al.}  [WMAP Collaboration],
  %``Wilkinson Microwave Anisotropy Probe (WMAP) three year results:
  %Implications for cosmology,''
  Astrophys.\ J.\ Suppl.\  {\bf 170}, 377 (2007)
  [arXiv:astro-ph/0603449].
  %%CITATION = APJSA,170,377;%%

\bibitem{LEP}
ALEPH Collaboration, DELPHI Collaboration, L3 Collaboration, OPAL Collaboration,
SLD Collaboration, LEP Electroweak Working Group Collaboration, SLD
Electroweak Group Collaboration, SLD Heavy Flavour Group Collaboration, Phys.
Rep. {\bf 427} (2006) 257.

 %\cite{Djouadi:2008uw}
\bibitem{Djouadi:2008uw}
  A.~Djouadi, M.~Drees, U.~Ellwanger, R.~Godbole, C.~Hugonie, S.~F.~King, S.~Lehti and S.~Moretti {\it et al.},
  %``Benchmark scenarios for the NMSSM,''
  JHEP {\bf 0807}, 002 (2008)
  [arXiv:0801.4321 [hep-ph]].
  %%CITATION = ARXIV:0801.4321;%%



\bibitem{ATLASupdate}
G.~Aad {\it et al.}  [ATLAS Collaboration], ATLAS-CONF-2012-168, 169, 170.

\bibitem{CMSupdate}
 S.~Chatrchyan {\it et al.}  [CMS Collaboration],
 CMS-PAS-HIG-12-045.


 %\cite{Ham:2007mt}
\bibitem{Ham:2007mt}
  S.~W.~Ham, S.~H.~Kim, S.~K.~OH and D.~Son,
  %``Higgs bosons of the NMSSM with explicit CP violation at the ILC,''
  Phys.\ Rev.\ D {\bf 76}, 115013 (2007)
  [arXiv:0708.2755 [hep-ph]];
  %%CITATION = ARXIV:0708.2755;%%
  %\cite{Ham:2008cg}
%\bibitem{Ham:2008cg}
  S.~W.~Ham, J.~O.~Im and S.~K.~OH,
  %``Neutral Higgs bosons in the MNMSSM with explicit CP violation,''
  Eur.\ Phys.\ J.\ C {\bf 58}, 579 (2008)
  [arXiv:0805.1115 [hep-ph]].
  %%CITATION = ARXIV:0805.1115;%%
  %\cite{King:2012tr}
%%\bibitem{King:2012tr}
%  S.~F.~King, M.~Muhlleitner, R.~Nevzorov and K.~Walz,
%  %``Natural NMSSM Higgs Bosons,''
%  arXiv:1211.5074 [hep-ph].
%  %%CITATION = ARXIV:1211.5074;%%

 %\cite{AlbornozVasquez:2011aa}
\bibitem{AlbornozVasquez:2011aa}
  D.~Albornoz Vasquez, G.~Belanger, R.~M.~Godbole and A.~Pukhov,
  %``The Higgs boson in the MSSM in light of the LHC,''
  Phys.\ Rev.\ D {\bf 85}, 115013 (2012)
  [arXiv:1112.2200 [hep-ph]].
  %%CITATION = ARXIV:1112.2200;%%

%%\cite{Robinett:1981yz}
%\bibitem{Robinett:1981yz}
%  R.~W.~Robinett and J.~L.~Rosner,
%  %``Prospects For A Second Neutral Vector Boson At Low Mass In SO(10),''
%  Phys.\ Rev.\  D {\bf 25}, 3036 (1982)
%  [Erratum-ibid.\  D {\bf 27}, 679 (1983)].
%  %%CITATION = PHRVA,D25,3036;%%
%
%
%%\cite{Langacker:1984dc}
%\bibitem{Langacker:1984dc}
%  P.~Langacker, R.~W.~Robinett and J.~L.~Rosner,
%  %``New Heavy Gauge Bosons In P P And P Anti-P Collisions,''
%  Phys.\ Rev.\  D {\bf 30}, 1470 (1984).
%  %%CITATION = PHRVA,D30,1470;%%



%%\cite{Suematsu:1998wm}
%\bibitem{Suematsu:1998wm}
%  D.~Suematsu,
%  %``Vacuum structure of the mu-problem solvable extra U(1) models,''
%  Phys.\ Rev.\  D {\bf 59} (1999) 055017
%  [arXiv:hep-ph/9808409].
%
%
%
%
%%\cite{Demir:2007dt}
%\bibitem{Demir:2007dt}
%  D.~A.~Demir, L.~L.~Everett and P.~Langacker,
%  %``Dirac Neutrino Masses from Generalized Supersymmetry Breaking,''
%  Phys.\ Rev.\ Lett.\  {\bf 100}, 091804 (2008)
%  [arXiv:0712.1341 [hep-ph]].
%  %%CITATION = PRLTA,100,091804;%%
%%\cite{AguilarSaavedra:2005pw}
%
%
%
%%\cite{Dai:1990xh}
%\bibitem{Dai:1990xh}
%  J.~Dai, H.~Dykstra, R.~G.~Leigh, S.~Paban and D.~Dicus,
%  %``CP VIOLATION FROM THREE GLUON OPERATORS IN THE SUPERSYMMETRIC STANDARD
%  %MODEL,''
%  Phys.\ Lett.\  B {\bf 237} (1990) 216
%  [Erratum-ibid.\  B {\bf 242} (1990) 547].



%
%
%  %\cite{Masip:1999mk}
%\bibitem{Masip:1999mk}
%  M.~Masip and A.~Pomarol,
%  %``Effects of SM Kaluza-Klein excitations on electroweak observables,''
%  Phys.\ Rev.\  D {\bf 60}, 096005 (1999)
%  [arXiv:hep-ph/9902467].
%  %%CITATION = PHRVA,D60,096005;%%
%
%%\cite{Suematsu:1997tv}
%\bibitem{Suematsu:1997tv}
%  D.~Suematsu,
%  %``Effect on the electron EDM due to abelian gauginos in SUSY extra U(1)
%  %models,''
%  Mod.\ Phys.\ Lett.\  A {\bf 12} (1997) 1709
%  [arXiv:hep-ph/9705412].


%%\cite{langackerx}
%\bibitem{langackerx}
%P.~Langacker and M.~Plumacher,
%  %``Flavor changing effects in theories with a heavy Z' boson with family
%  %non-universal couplings,''
%  Phys.\ Rev.\  D {\bf 62}, 013006 (2000)
%  [arXiv:hep-ph/0001204].
%  %%CITATION = PHRVA,D62,013006;%%
%
% %\cite{correlate}
%  \bibitem{correlate}
%  T.~M.~Aliev, D.~A.~Demir, E.~Iltan and N.~K.~Pak,
%  %``The CP Asymmetry in $b \to s l~+ l~-$ Decay,''
%  Phys.\ Rev.\  D {\bf 54}, 851 (1996)
%  [arXiv:hep-ph/9511352].
%  %%CITATION = PHRVA,D54,851;%%
%
%%\cite{correlate2}
%\bibitem{correlate2}
%D.~A.~Demir,
%  %``Higgs boson couplings to quarks with supersymmetric CP and flavor
%  %violations,''
%  Phys.\ Lett.\  B {\bf 571}, 193 (2003)
%  [arXiv:hep-ph/0303249];
%  %%CITATION = PHLTA,B571,193;%%
%A.~Dedes and A.~Pilaftsis,
%  %``Resummed effective Lagrangian for Higgs-mediated FCNC interactions in the
%  %CP-violating MSSM,''
%  Phys.\ Rev.\  D {\bf 67}, 015012 (2003)
%  [arXiv:hep-ph/0209306];
%  %%CITATION = PHRVA,D67,015012;%%
%M.~S.~Carena, A.~Menon, R.~Noriega-Papaqui, A.~Szynkman and
%C.~E.~M.~Wagner,
%  %``Constraints on B and Higgs physics in minimal low energy supersymmetric
%  %models,''
%  Phys.\ Rev.\  D {\bf 74}, 015009 (2006)
%  [arXiv:hep-ph/0603106].
%  %%CITATION = PHRVA,D74,015009;%%
%



\end{thebibliography}
\end{document}